\documentstyle[12pt,epsfig,ldd_art]{article} 
\input{epsf.tex}
\input{psfig.sty}
\newcommand{\beq}{\begin{equation}}
\newcommand{\eeq}{\end{equation}}
\newcommand{\bea}{\begin{eqnarray}}
\newcommand{\eea}{\end{eqnarray}}

\newcommand{\cond}{\langle\bar\chi\chi\rangle}

\newcommand{\oh}{\frac{1}{2}}

\newcommand{\non}{\nonumber}

\newcommand{\de}[1]{\frac{\partial}{\partial #1}}

\newcommand{\One}{1\kern-4.5pt1}
\begin{document}

\addtolength{\baselineskip}{0.20\baselineskip}

\hfill hep-lat/9701016

\hfill SWAT/97/136

\hfill January 1997

\begin{center}

\vspace{48pt}

{ {\bf The Three Dimensional Thirring Model for Small $N_f$} }

\end{center}

\vspace{18pt}

\centerline{\sl L. Del Debbio\footnote{present address:
Centre de Physique Th\'eorique, CNRS Luminy, Case 907,
F-13288 Marseille Cedex 9, France}, S.J. Hands, J.C. Mehegan}
\centerline{\sl (the UKQCD collaboration)}

\vspace{20pt}


\centerline{Department of Physics, University of Wales Swansea,}
\centerline{Singleton Park, Swansea SA2 8PP, U.K.}

\vspace{48pt}

\begin{center}  

{\bf Abstract}

\end{center}

We formulate  the three dimensional Thirring
model on a spacetime lattice and study it for various 
even numbers of fermion flavors $N_f$ by Monte Carlo simulation. We find
clear evidence for
spontaneous chiral symmetry breaking at strong coupling, contradicting
the predictions of the $1/N_f$ expansion.
The critical point appears to correspond to an ultra-violet fixed
point of the renormalisation group; a fit to a RG-inspired equation of
state in the vicinity of the fixed point yields distinct critical exponents
for $N_f=2$ and $N_f=4$, while no fit is found for $N_f=6$, suggesting
there is a critical number $N_{fc}<6$ beyond which no chiral symmetry 
breaking occurs. 
The spectrum of the $N_f=2$ theory is studied; the states examined
vary sharply but continuously across the transition.



\vfill

\newpage

\section{Introduction}

The Thirring model is a quantum field theory of fermions interacting
via a current-current contact term, described in three dimensional
continuum Euclidean
spacetime by the following Lagrangian:
\begin{equation}
	{\cal L}=\bar\psi_i(\partial{\!\!\! /}\,+m)\psi_i
	+{g^2\over{2N_f}}(\bar\psi_i\gamma_\mu\psi_i)^2.
\label{eq:thirring}
\end{equation}
Here $\psi$ and $\bar\psi$ are four-component spinors, $m$ is a bare mass, 
and $i$ runs over $N_f$ fermion species. After the introduction of an 
auxiliary vector field, the above Lagrangian can be rewritten as:
\begin{equation}
	{\cal L}_{aux} = \bar\psi_i(\partial{\!\!\! /}\,
	+ iA{\!\!\! /}\, + m)\psi_i 
	+ {N_f\over{2g^2}} (A_\mu)^2
\label{eq:auxiliary}
\end{equation}
which now contains the fermionic bilinear term of the usual QED lagrangian
but with  a mass term for the vector field, which   
spoils the gauge-invariance of ${\cal L}_{aux}$. This is no surprise, since
the original lagrangian did not show any local symmetry. The global
symmetries are a $U(N_f)$ one, corresponding to rotations in flavor 
space, and, for $m\rightarrow 0$, a $U(N_f)_A$ chiral symmetry. 
Since the coupling constant $g$ has
mass dimension $[g]=(1-{d\over2})=-{1\over2}$, 
the usual perturbative expansion is
non-renormalisable. However perturbation theory is not the end of the 
story and, as suggested several years ago~\cite{Parisi75,Gomes91}, different
approaches can yield a well-defined continuum limit, 
corresponding to a UV--stable fixed point of the renormalisation group (RG).

\begin{figure}[htbp]
\vspace{3pt}
\centerline{
\setlength\epsfxsize{300pt}
\epsfbox{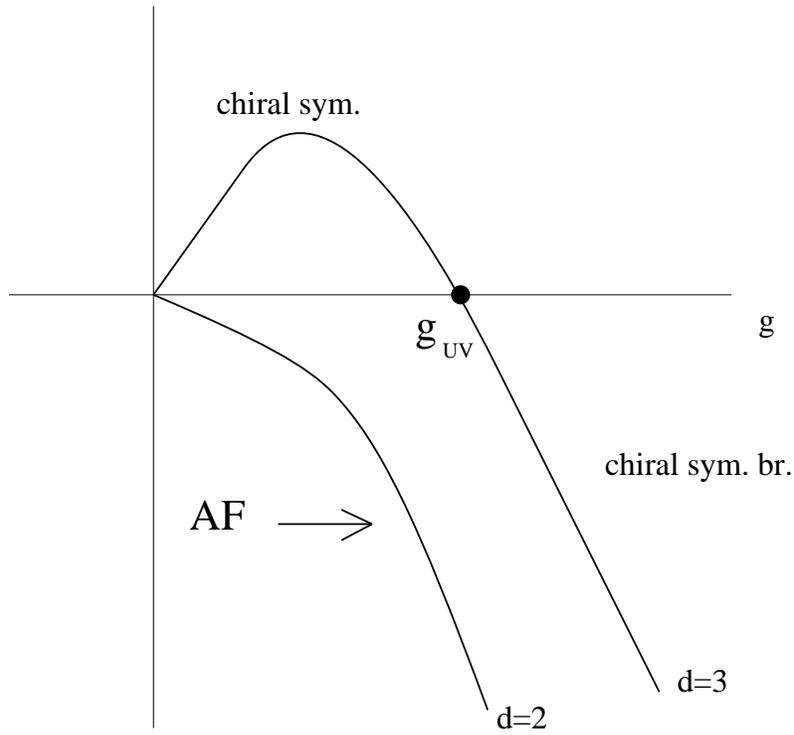}}
\caption{$\beta$-function for $d=2$ and $2<d<4$, showing for the
latter case a non-trivial UV fixed point and 
phase structure.
\label{fig:beta_function}}
\end{figure}

The Gross--Neveu model (GN), which has been studied both numerically and 
in the $1/N_f$ expansion~\cite{Hands93,Rosenstein91,Bielefeld}, exhibits 
this behaviour.  The model turns out to be renormalisable to all orders
in $1/N_f$, with a UV--stable RG fixed point at $g_{UV} \neq 0$ and chiral 
symmetry breaking for $g > g_{UV}$; only for $g\simeq g_{UV}$
can the ratio of the physical scale (either a particle mass or a 
scattering cross-section) to the UV cutoff be made small. The theory 
defined at this point describes the IR dynamics of a linear sigma
model with a Yukawa coupling
to fermions~\cite{Zinn-Justin}, which is super-renormalisable about the origin:
\beq
	{\cal L}_{Yu} = \bar\psi (\partial{\!\!\! /}\, + v \sigma +
	m) \psi + \oh(\partial \sigma)^2 + \oh \mu^2 \sigma^2 +
	\lambda \sigma^4
\eeq
For a critical value of $v$ in the IR regime the kinetic and 
quartic terms of the $\sigma$-field
become irrelevant operators and thus the system equivalent to the GN model.
The critical behaviour for the GN model can be summarised by the behaviour 
of the $\beta$-function, as shown in Fig.~\ref{fig:beta_function}. In two
dimensions, the theory is asymptotically free, and the $\beta$-function
exhibits the usual UV fixed point at the origin. For $2<d<4$, the 
fixed point is shifted away from the origin, separating a strong coupling
regime, where dynamical chiral symmetry breaking is possible, from a
weak coupling phase where the theory is chirally 
symmetric~\cite{Muta92}.

The situation for the Thirring model is less clear.
At both leading and next-to-leading order in $1/N_f$ the 
renormalisability of the massless model
has been established~\cite{Hands95}. For $d<4$ the vacuum polarisation
is actually UV--finite so long as the regularisation respects
current conservation. As a consequence a continuum limit may be defined
for {\sl any} value of the coupling $g$, which would control, e.g., the
ratio of physical fermion mass to vector bound state mass. In other words,
the $\beta$-function is vanishing for any value of $g$ and the interaction
$(\bar\psi_i\gamma_\mu\psi_i)^2$ is a {\sl marginal} operator, whereas the 
interaction for the GN model is {\sl relevant}~\cite{Gat92}.
Another prediction of the $1/N_f$ expansion is the vanishing of the 
vector bound
state mass in the strong coupling limit $mg \rightarrow \infty$. A 
massless particle mediating an interaction between conserved currents
suggests an equivalence between the UV behaviour of Thirring model and the
IR one of $\mbox{QED}_3$, which would 
thus be the analogue of the equivalence between the
GN and Yukawa models. Such a conjecture is supported by the fact that the 
$O(1/N_f)$ corrections in the two models appear to 
coincide~\cite{Hands95,Espriu82}.
Moreover
in the $1/N_f$ expansion, as in perturbative gauge theory, no diagram
ever appears resulting in the dynamical breaking of chiral symmetry leading
to spontaneous mass generation.

The prediction of no spontaneous mass generation is contradicted, however, 
by results from a completely
non-perturbative approach, namely solution of Schwinger--Dyson (SD) equations,
which predicts a non-vanishing chiral condensate $\langle\bar\psi\psi\rangle$
in the limit $m \rightarrow 0$~\cite{Gomes91,Hong94,Itoh95}. In this approach a
non-trivial solution for the dressed fermion propagator
$S(p)=(A(p)ip{\!\!\! /}\,+\Sigma(p))^{-1}$ is sought, i.e. one in which
the self-energy $\Sigma(p)$ and hence $\langle\bar\psi\psi\rangle$ are
non-vanishing in the chiral limit $m\to0$. Unfortunately, the SD
equations can only be solved by truncating them in a somewhat arbitrary
fashion. The usual approximation~\cite{Gomes91,Itoh95}, is to assume that the
vector propagator is given by its leading-order form for $m=0$ in the
$1/N_f$ expansion, viz.
\beq
D_{\mu\nu}(k)=\left(\delta_{\mu\nu}-{{k_\mu k_\nu}\over k^2}\right)
\left(1+{{g^2(k^2)^{1\over2}}\over8}\right)^{-1}
+{{k_\mu k_\nu}\over k^2}
\label{eq:Dmunu}
\eeq
and that the fermion-vector vertex function is well-approximated by the
bare vertex (the so-called ``planar'' or ``ladder'' approximation):
\beq
\Gamma_\mu(p,q)=-{{ig}\over\sqrt{N_f}}\gamma_\mu.
\eeq

The most systematic treatment has been given by Itoh et al \cite{Itoh95}, 
who note the equivalence of (\ref{eq:auxiliary})
to a gauge-fixed form of a gauged fermion-scalar model. This allows the 
specification of an alternative non-local gauge-fixing condition, which in 
turn entails $A(p)\equiv1$, with no effect, of course, on gauge-invariant 
quantities such as the physical mass
or $\langle\bar\psi\psi\rangle$. The resulting SD equations can be 
solved exactly in the limit $g^2\to\infty$; a non-trivial solution
for $\Sigma(p)$ exists for
\beq
N_f<N_{fc}={128\over{3\pi^2}}\simeq4.32.
\eeq
This is identical to the critical $N_{fc}$ predicted for non-trivial IR 
behaviour in $\mbox{QED}_3$ \cite{QED3}.
However, since the integral equations require the introduction of a UV
cutoff $\Lambda$, a feature of this solution is that the induced physical 
scale $\mu$ depends on $N_f$ in an essentially singular way:
\beq
{\mu\over\Lambda}\propto\exp\left(-{2\pi\over\sqrt{{N_{fc}\over
N_f}-1}}\right);\;\;\;
\langle\bar\psi\psi\rangle\propto\Lambda^{1\over2}\mu^{3\over2}\propto
\exp\left(-{3\pi\over\sqrt{{N_{fc}\over
N_f}-1}}\right)
\label{eq:nonanal}
\eeq
This implies that in the strong coupling limit 
a continuum limit only exists as $N_f\to N_{fc}$. 

The behaviour (\ref{eq:nonanal}) is an example of an infinite order, or 
``conformal'' phase transition, originally discussed in a particle physics 
context by Miranskii and collaborators in quenched $\mbox{QED}_4$ 
\cite{Miranskii83}, and recently discussed in more generality \cite{MirYam96}.
One prediction of 
\cite{MirYam96} is that although the order parameter 
$\langle\bar\psi\psi\rangle$ is continuous at the transition, there is a 
discontinuity in the spectrum of light excitations, with no light scalar 
resonances on the symmetric side of the transition. In the broken phase, of 
course, Goldstone's theorem predicts light pions.
This behaviour has recently been exhibited in an investigation of
four dimensional SU($N_c$) gauge theory with an intermediate number $N_f$ of 
fermion flavors \cite{Appel96}: a solution of the Bethe-Salpeter equation in 
ladder approximation found no low-momentum pole in the symmetric phase.

Unfortunately no
analytic solution to the SD equation for the Thirring model exists 
for $g^2<\infty$; however using different
techniques Kondo~\cite{Kondo95} has argued that there is a critical 
line $N_{fc}(g^2)$ in the $(g^2,N_f)$ plane, which is a smooth invertible 
function.
Therefore for integer $N_f<N_{fc}$ one might conjecture the form
\cite{ldd96}
\beq
\langle\bar\psi\psi\rangle\propto\exp\left(-{a\over\sqrt{{g^2\over
g_c^2}-1}}\right),
\label{eq:conformal}
\eeq
corresponding to a symmetry restoring transition at some critical point
$g^2=g_c^2(N_f)$. Presumably in this scenario the coupling $g$ has
become {\sl relevant\/}: there may exist a novel strongly-coupled continuum
limit at the critical point not described by the $1/N_f$ expansion,
which is simultaneously the UV limit of the Thirring model and the IR limit 
of $\mbox{QED}_3$. Since there is no small dimensionless parameter in play, 
this fixed point would be inherently non-perturbative. 
It is worth mentioning in passing
that a novel IR fixed point in $\mbox{QED}_3$ has been proposed by Aitchison and
Mavromatos in connection with non-Fermi liquid behaviour in the normal phase
of superconducting cuprates \cite{Aitch95}.

Although this scenario of a non-trivial fixed point for $N_f<N_{fc}$
is attractive, 
there are good reasons to be cautious. Using a
different sequence of truncations Hong and Park have found chiral
symmetry  breaking for {\sl all\/} $N_f$~\cite{Hong94}, with
\beq
{1\over g_c^2}\propto\exp\left(-{{N_f\pi^2}\over16}\right),
\label{eq:HongPark}
\eeq
a result which is also non-analytic in $1/N_f$ and hence beyond the 
reach of the 
$1/N_f$ perturbative series. Moreover in  $\mbox{QED}_3$,  studies beyond the
planar approximation, using improved ans\"atze for the vertex
$\Gamma_\mu$, suggest that the condition $A(p)\equiv1$ is unphysical,
and that chiral symmetry is spontaneously broken for all $N_f$
~\cite{Pennington88}.
In the current context this would imply $g_c^2<\infty$ for all $N_f$.

Because of the systematic uncertainty attached to the SD approach, and because 
of the intrinsic interest in exploring the non-perturbative behaviour of a 
fermionic model, we have studied the three dimensional Thirring model by 
Monte Carlo simulation on a spacetime lattice. Our initial results appeared 
in \cite{ldd96}, where we claimed that critical points with distinct critical 
indices exist for the models with $N_f=2,4$, but that evidence for a 
critical theory with $N_f=6$ was not found.
In this paper we summarise those results, and present further data 
which enable us to refine our description of the critical behaviour of the 
$N_f=2$ model. In Sec. 2 we give a fairly lengthy exposition of the lattice 
formulation, showing that the lattice method is itself not without 
systematic uncertainties. Although we use staggered lattice fermions which 
have a continuous remnant of the $\mbox{U}(N_f)_A$ axial symmetry, a 
transcription to a basis in which the Fermi fields have explicit spin and 
flavor indices reveals that there are many other interactions present at 
tree level besides the desired current-current form. Perhaps a robust 
counter-argument to this problem is that 
for a strongly-coupled  fixed point a non-perturbative regularisation is 
always necessary, and therefore ``what you see is what you get'', i.e. the 
fixed point theory defines itself via its critical exponents and spectral 
quantities.
In addition, the lattice form of the current is not exactly conserved, 
resulting in an additive renormalisation of the inverse coupling $1/g^2$. 
This makes it difficult in practice to identify the strong coupling limit 
$g^2\to\infty$.  A description of the simulation, and a list of the measured 
observables, namely the condensate, susceptibilities and propagators, follows.

In Sec. 3 we present results for the equation of state, i.e.
$m=f(\langle\bar\psi\psi\rangle,g)$. Since it is impossible to work directly 
in the chiral limit $m=0$, we find it convenient to fit to a phenomenological 
form inspired by the standard power-law scaling form used in the theory of 
ferromagnetism, with its attendant critical exponents $\beta,\gamma,\delta$ 
and $\nu$, together with the additional constraints of hyperscaling, and, from 
the ladder approach to the gauged Nambu -- Jona-Lasinio model \cite{Kocic91}, 
that $\delta-1/\beta=\gamma\equiv1$. We justify this approach not so much by 
any theoretical argument but because its shortcomings if any should manifest 
themselves as variations in the fitted values of the exponents as data is 
taken from systems on larger volumes and closer to the chiral limit. In
\cite{ldd96} the fitted exponents for the $N_f=2,4$ models were distinct, a 
feature not seen in similar studies of $\mbox{QED}_4$ \cite{KKW93,QED4}. Here 
we supplement the data of \cite{ldd96} for $N_f=2$ with results from a larger 
lattice $(16^3)$ and a smaller bare mass $m$, and refine the fit by combining 
data from several lattice volumes in a finite size scaling analysis. We find 
critical exponents and $g_c^2$ completely consistent with those of our earlier 
studies, which strengthens the case for a critical point described by 
power-law scaling. We also have some new results for the $N_f=6$ model on 
$12^3$ systems,
supporting our previous claim that there is no transition in this case, 
and hence that $N_{fc}<6$ for the Thirring model.

In Sec. 4 we present exploratory results for susceptibilities, and for 
fermion and bound-state masses. Our lattice size of $16^3$ is unfortunately 
not long enough in the temporal direction to permit real accuracy; nonetheless 
qualitative features of the level ordering in either phase do emerge. The 
susceptibilities reflect the phase transition very nicely, with longitudinal 
and transverse susceptibilities $\chi_l$ and $\chi_t$ varying from 
near-degeneracy  in the symmetric phase (thus indirectly supporting the 
hypothesis $\delta-1/\beta=1$) to $\chi_t\gg\chi_l$ in the broken phase. The 
effects of dynamical fermion loops in the quantum vacuum can also be 
estimated, and appear to become dominant in the broken phase. The fermion, 
pseudoscalar and scalar masses all behave as expected, the pion mass staying 
small and roughly constant, while fermion and scalar masses change sharply 
but continuously across the transition. There is no sign of the 
discontinuities described in \cite{MirYam96}. Finally studies in the 
vector-like channel reveal light states both with vector and axial vector 
quantum numbers in the symmetric phase. This is interesting because no light 
axial vector is predicted in the continuum $1/N_f$ expansion 
at leading order, as we show in Sec.~4.
Both states become difficult to observe in the broken phase. Our conclusions 
are presented in Sec. 5.

\section{Lattice formulation}

We want to discretise the action of Eq.~(\ref{eq:thirring}) on a hypercubic
Euclidean lattice. In order to recover a partial $U(N)_A$ axial symmetry
in the massless limit, we use $N$ flavors of staggered fermions.
The lattice action then reads:
\bea
	S &=& \oh \sum_{x\mu i}
 \bar\chi_i(x) \eta_\mu(x) \left[\chi_i(x+\hat\mu) -
	\chi_i(x-\hat\mu)\right] + \non \\
	  & & m \sum_{xi} \bar\chi_i(x) \chi_i(x) + 
\frac{g^2}{2N} \sum_{x\mu ij} \bar\chi_i(x) \chi_i(x+\hat\mu) 
	\bar\chi_j(x+\hat\mu) \chi_j(x)
\label{eq:lattice-thirring}
\eea
where $\chi, \bar\chi$ are the staggered fermion fields, $\eta_\mu$ the
Kawamoto--Smit phases, $m$ is the bare fermion mass, 
the flavor indices $i,j$ run from
$1$ to $N$ and the vector character of the interaction is encoded in the 
geometrical structure of the four fermion term.

To express the action as a bilinear in $\chi,\bar\chi$, which makes it
amenable to simulation, we introduce a bosonic auxiliary field
as in (\ref{eq:auxiliary}). In this
paper we choose a form which we will refer to as {\sl non-compact\/}:
\bea
	S &=& \oh \sum_{x\mu i} \bar\chi_i(x) \eta_\mu(x) 
	(1+iA_\mu(x)) \chi_i(x+\hat\mu)
	+ \mbox{h.c.} \non \\
	  & & + m \sum_{xi} \bar\chi_i(x) \chi_i(x) +
	\frac{N}{4g^2} \sum_{x\mu} A_\mu(x)^2 \non \\
	  &\equiv& \sum_{xyi}\bar\chi_i(x) M[A,m](x,y) \chi_i(y) + 
          \frac{N}{4g^2} \sum_{x\mu} A_\mu(x)^2 
\label{eq:non-compact}
\eea
where we have introduced $M[A,m]$ to denote the
fermionic bilinear, which depends on both the auxiliary field and the 
bare mass.
It is straightforward to integrate over the auxiliary field 
and recover Eq.~(\ref{eq:lattice-thirring}). 

In the chiral limit $m\to0$ both the original action
(\ref{eq:lattice-thirring}) and the bosonised form
(\ref{eq:non-compact}) are invariant under a global
axial rotation, which for $N=1$ is written
\bea
\chi(x)\mapsto\exp(i\alpha\varepsilon(x))\chi(x)\;\;\;;\;\;\;
\bar\chi(x)\mapsto\exp(i\alpha\varepsilon(x))\bar\chi(x),
\label{eq:chsymmetry}
\eea
where the phase $\varepsilon(x)=(-1)^{x_1+x_2+x_3}$.
The generalisation to U(N) is straightforward.
We will refer to this as the {\sl chiral symmetry\/} of the model. In the
$m\to0$ limit for certain values of the coupling $g^2>g_c^2(N)$ it is
spontaneously broken, signalled by the appearance of a condensate
$\langle\bar\chi\chi\rangle\not=0$. 

Some comments on the lattice formulation are appropriate. 
Eq. (\ref{eq:non-compact}) in the long wavelength limit
(i.e. assuming $a\partial\psi\ll\psi$) is clearly closely
related to the continuum auxiliary form (\ref{eq:auxiliary}), and as 
such is suitable for a perturbative
expansion in small-mode fluctuations of the $A_\mu$ fields, corresponding
to the continuum $1/N_f$ expansion. This expansion is well-behaved
because to leading order the vector propagator $D_{\mu\nu}(x)=\langle
A_\mu(0)A_\nu(x)\rangle$ receives a contribution not just from the
quadratic term $A_\mu^2$, but also from the vacuum polarisation loop 
$\Pi_{\mu\nu}$, which is finite in $d<4$. We find in the deep Euclidean
region~\cite{Hands95} (Cf. (\ref{eq:Dmunu}))
\begin{equation}
\displaystyle\lim_{k^2\to\infty}D_{\mu\nu}(k)\propto{\cal P}_{\mu\nu}(k)
{1\over\sqrt{k^2}}+\mbox{longitudinal}
\end{equation}
where ${\cal P}_{\mu\nu}$ is the transverse projector
\bea
{\cal P}_{\mu\nu}(k)=\delta_{\mu\nu}-{{k_\mu k_\nu}\over k^2}.
\eea
Because the interaction current $\bar\psi\gamma_\mu\psi$ is conserved,
the longitudinal component of $D_{\mu\nu}$ has no physical 
significance~\cite{Parisi75}\cite{Itoh95}\cite{Kondo95}. 
The $1/k$ behaviour of $D_{\mu\nu}$ in
effect suppresses short wavelength excitations of the auxiliary field,
and justifies the long-wavelength expansion {\sl a posteriori\/}. It 
means we might hope the form (\ref{eq:non-compact}) has the correct
continuum limit. This assumption may not be justified, however, 
due to the chiral phase transition 
at finite $g^2$. Since the $1/N_f$ expansion predicts
no phase transition, we know that there will be at least some part of
coupling constant space where the above arguments cannot hold. In
this case it is not quite so clear that (\ref{eq:non-compact})
coincides with the continuum form (\ref{eq:thirring}).

To see why there might be a problem, we must consider the continuum
flavor interpretation of the lattice action (\ref{eq:lattice-thirring}).
First define a unitary transformation to fields $q,\bar q$~\cite{BB}:
\bea
q^{\alpha a}_i(y)=\left(\matrix{u^{\alpha a}_i(y)\cr
                                d^{\alpha a}_i(y)\cr}\right)
={1\over{4\surd{2}}}\sum_P\left(\matrix{
\Gamma^{\alpha a}_P&\cr &B^{\alpha a}_P\cr}\right)\chi_i(P;y),
\label{eq:b+b}
\eea
where $y$ denotes a site on a lattice of spacing $2a$ and $P$ is a
3-vector with entries either 0 or 1 ranging over corners of the
elementary cube associated with $y$, so that each side of the original
lattice corresponds to a unique choice of $(P;y)$. The $2\times2$
matrices $\Gamma_P$ and $B_P$ are defined by
\bea
\Gamma_P=\tau_1^{P_1}\tau_2^{P_2}\tau_3^{P_3}\;\;\;;\;\;\;
B_P=(-\tau_1)^{P_1}(-\tau_2)^{P_2}(-\tau_3)^{P_3},
\eea
where $\tau_\mu$ are the Pauli matrices.
It is then possible to recast the bilinear term of
(\ref{eq:lattice-thirring}) as follows;
\bea
S_{kin} &=& (2a)^3\sum_{y\mu}\biggl( 
\bar q_i(y)(\gamma_\mu\otimes\One_2){[q_i(y+\hat\mu)-q_i(y-\hat\mu)]\over
{4a}}+\non\\
& &\bar q_i(y)(\gamma_4\otimes\tau_\mu^*)
{[q_i(y+\hat\mu)+q_i(y-\hat\mu)-2q_i(y)]\over{4a}}\biggr)+\non\\
& &m(2a)^3\sum_y \bar q_i(y)(\One_4\otimes\One_2)q_i(y).
\label{eq:cont-kin}
\eea
The direct product is between $4\times4$ Dirac matrices, acting on
spinor degrees of freedom, defined by
\bea
\gamma_\mu=\left(\matrix{\tau_\mu&\cr&-\tau_\mu\cr}\right)\;\;\;;\;\;\;
\gamma_4=\left(\matrix{&-i\One_2\cr i\One_2&\cr}\right)\;\;\;;\;\;\;
\gamma_5=\left(\matrix{&\One_2\cr\One_2&\cr}\right),
\eea
and $2\times2$ matrices acting on flavor degrees of freedom. Each
lattice fermion flavor in three dimensions corresponds to
two continuum four-component spinor flavors. Hence
\bea
N_f=2N.
\label{eq:N_f=2N}
\eea
In terms of this new basis the chiral symmetry of the model
(\ref{eq:chsymmetry}) reads
\bea
q\mapsto\exp(i\alpha(\gamma_5\otimes\One))q\;\;\;;\;\;\;
\bar q\mapsto\bar q\exp(i\alpha(\gamma_5\otimes\One)).
\eea
The condensate $\langle\bar\chi\chi\rangle$ in the new basis 
reads $\langle\bar q(\One\otimes\One)q\rangle$.

The first and third terms of (\ref{eq:cont-kin}) are clearly of the
same form as the continuum action for free massive fermions, whereas the second
term, which is flavor non-singlet and Lorentz non-covariant, is
formally $O(a)$ and hence hopefully irrelevant. The problem comes when
we consider the interaction term. In the auxiliary field formalism, 
only in the long-wavelength limit would we expect the interaction to
assume its expected continuum form $A_\mu\bar
q_i(\gamma_\mu\otimes\One_2)q_i$~\cite{GoltSmit}.
When $A$ is integrated out to recover the original form
(\ref{eq:lattice-thirring}) though, all momentum modes are summed over
with equal weight. In fact, in four-fermi form the interaction
contains many extra terms\footnote{The corresponding result
in four dimensions was first communicated 
to us by M. G\"ockeler~\cite{Gockpc}}:
\bea
S_{int}&=&{g^2\over{2N}}\times{(2a)^3\over2}\times\sum_{y\mu ij}\non\\
& &\biggl[\Bigl(\bar q_i(y)(\gamma_\mu\otimes\One)(1+f_\mu)q_i(y)
+\bar q_i(y)(i\gamma_4\otimes\tau_\mu^*)\delta_\mu
q_i(y)\Bigr)\times\non\\
& &\Bigl((1+f_\mu)\bar q_j(y)(\gamma_\mu\otimes\One)q_j(y)+
\delta_\mu\bar q_j(y)(-i\gamma_4\otimes\tau_\mu^*)q_j(y)\Bigr)\non\\
&+&\Bigl(\bar q_i(y)(\gamma_5\gamma_\mu\otimes\One)(1+f_\mu)q_i(y)+
\bar q_i(y)(i\gamma_5\gamma_4\otimes\tau_\mu^*)\delta_\mu
q_i(y)\Bigr)\times\non\\
& &\Bigl((1+f_\mu)\bar q_j(y)(\gamma_\mu\gamma_5\otimes\One)q_j(y)+
\delta_\mu\bar q_j(y)(-i\gamma_4\gamma_5\otimes\tau_\mu^*)q_j(y)\Bigr)\non\\
&+&\sum_\nu\Bigl(\bar
q_i(y)(i\gamma_4\gamma_5\gamma_\nu\gamma_\mu\otimes\tau_\nu^*)(1+f_\mu)
q_i(y)+\bar q_i(y)(\gamma_\nu\gamma_5\otimes\tau_\nu^*\tau_\mu^*)
\delta_\mu q_i(y)\Bigr)\times\non\\
& &\Bigl((1+f_\mu)\bar q_j(y)(i\gamma_4\gamma_5\gamma_\mu\gamma_\nu
\otimes\tau_\nu^*)q_j(y)+\delta_\mu\bar q_j(y)(-\gamma_\nu\gamma_5\otimes
\tau_\mu^*\tau_\nu^*)q_j(y)\Bigr)\non\\
&+&\sum_\nu\Bigl(\bar
q_i(y)(-i\gamma_4\gamma_\nu\gamma_\mu\otimes\tau_\nu^*)(1+f_\mu)
q_i(y)+\bar q_i(y)(-\gamma_\nu\otimes\tau_\nu^*\tau_\mu^*)
\delta_\mu q_i(y)\Bigr)\times\non\\
& &\Bigl((1+f_\mu)\bar q_j(y)(i\gamma_4\gamma_\mu\gamma_\nu
\otimes\tau_\nu^*)q_j(y)+\delta_\mu\bar q_j(y)(-\gamma_\nu\otimes
\tau_\mu^*\tau_\nu^*)q_j(y)\Bigr)\biggr],
\label{eq:OhGod}
\eea
where the difference operators $f$ and $\delta$ are defined by
\bea
f_\mu q(y)={1\over2}\left(q(y+\hat\mu)+q(y-\hat\mu)\right)\;\;\;;\;\;\;
\delta_\mu q(y)={1\over2}\left(q(y+\hat\mu)-q(y-\hat\mu)\right).
\eea
The terms containing $\delta$ are formally $O(a)$ and hence probably
irrelevant, but no such argument can be used for the remainder, which,
in addition to the desired form $(\bar q_i(\gamma_\mu\otimes\One)q_i)^2$,
contains extra flavor non-singlet and non-covariant terms. If 
the $1/N_f$ expansion were valid, then one could argue that only
interactions between conserved currents would generate long-range
correlations between the $A$ fields, and that all other interactions 
would become irrelevant in the continuum limit -- a similar
effect appears to be
the case in other two~\cite{CER,Joli}
and three~\cite{Hands93} dimensional four-fermi models where a renormalisable
expansion is available. Even in this case there might also be
contributions in the chiral limit from the interaction
$(\bar q_i(\gamma_\mu\gamma_5\otimes\One)q_i)^2$. However, as argued above,
in the present case we know the $1/N_f$ expansion fails qualitatively
in the chirally broken phase. It therefore becomes a matter of
experiment, by studying the current-current and $AA$ correlations in
various channels numerically, to determine the continuum limit of the
action (\ref{eq:lattice-thirring}).

\begin{figure}[htbp]
\vspace{3pt}
\centerline{
\setlength\epsfxsize{300pt}
\epsfbox{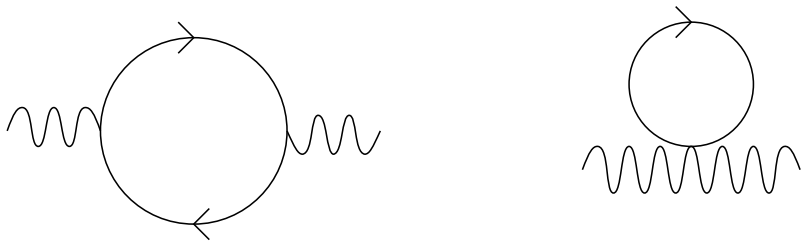}}
\caption{Diagrams contributing to vacuum polarisation in lattice QED.
\label{fig:diagrams}}
\end{figure}

A second issue we must address is that despite the above comments there
is an important difference between the continuum $1/N_f$ expansion and
the $1/N$ expansion applied to the non-compact lattice action
(\ref{eq:non-compact}). This concerns vector current conservation. In
the continuum model a regularisation such as Pauli-Villars, which
conserves the interaction current, must be assumed in order to ensure
that the superficial $O(\Lambda)$ contribution to vacuum polarisation
vanishes, leaving a finite result~\cite{Parisi75,Gomes91,Hands95}. 
Now, in non-compact lattice $\mbox{QED}_3$, the vacuum
polarisation is again UV finite, but this time because of a cancellation
between $O(a^{-1})$ contributions from two separate
diagrams (Fig.~\ref{fig:diagrams} - see eg.~\cite{Rothe}).
The second diagram appears because the gauge-invariant interaction term
in QED is $\bar\chi\exp(ieA_\mu)\chi$, and hence contains
$n$-photon-electron-positron vertices. In the action
(\ref{eq:non-compact}), on the other hand, the interaction is 
$\bar\chi iA_\mu\chi$, so only the left-hand diagram is present.
Therefore cancellation of $O(a^{-1})$ divergences does not occur. We
find for the vacuum polarisation tensor to leading order in $1/N$:
\bea
\displaystyle\lim_{k^2a^2\to0}
\Pi_{\mu\nu}^{LATT}(k)={\cal P}_{\mu\nu}(k)\Pi^{CONT}(k^2,m)+{g^2\over
a}
\delta_{\mu\nu}J(m),
\eea
where 
\bea
J(m)=2\int_{-\pi}^\pi{{d^3q}\over(2\pi)^3}{{\sin^2q_\mu}\over
{\sum_\nu\sin^2q_\nu+m^2}}\;\;\;;\;\;\;
\displaystyle\lim_{m\to0}J(m)={2\over3},
\eea
and, recalling a factor of two from
(\ref{eq:N_f=2N}) 
\bea
\Pi^{CONT}(k^2,m)=-{g^2\over\pi}\left[
m+{(k^2-4m^2)\over{2\sqrt{k^2}}}\tan^{-1}\left({\sqrt{k^2}\over{2m}}\right)
\right].
\eea
This means that for the inverse auxiliary propagator (setting $a=1$)
\bea
D_{\mu\nu}^{-1}(k)&=&\delta_{\mu\nu}-\Pi^{LATT}_{\mu\nu}(k)\non\\
&=&\delta_{\mu\nu}(1-g^2J(m))-{\cal P}_{\mu\nu}(k)\Pi^{CONT}(k^2,m)
\label{eq:Dmunu-1}
\eea
i.e.
\bea
D^{LATT}_{\mu\nu}(k)={1\over{1-g^2J(m)}}\left[{{{\cal P}_{\mu\nu}(k)}
\over{1-{g^2\over{1-g^2J(m)}}\Pi^{CONT}(k^2,m)}}+{{k_\mu
k_\nu}\over{k^2}}\right].
\eea
If we compare this with the continuum result
\bea
D^{CONT}_{\mu\nu}(k)={{{\cal
P}_{\mu\nu}(k)}\over{1-g^2\Pi^{CONT}(k^2,m)}}+{{k_\mu k_\nu}\over k^2},
\eea
we see that the $O(a^{-1})$ divergence can be absorbed by a
wavefunction renormalisation together with a coupling constant
renormalisation:
\bea
A\mapsto A_R=(1-g^2J(m))^{1\over2}A\;\;\;;\;\;\;
g^2\mapsto g^2_R={g^2\over{1-g^2J(m)}}.
\eea

Now, the physics described by continuum $1/N_f$ perturbation theory
occurs for the range of couplings $g^2_R\in[0,\infty)$, i.e. for
$g^2\in[0,g^2_{lim})$, where to leading order
\bea
\displaystyle\lim_{m\to0}{1\over g^2_{lim}}={2\over3}.
\label{eq:glim}
\eea
We therefore expect discontinuous behaviour at bare coupling
$g^2\simeq g^2_{lim}$, since at this point the auxiliary propagator is
not defined (i.e. $D_{\mu\nu}^{-1}$ in (\ref{eq:Dmunu-1}) has no
inverse). For $g^2>g^2_{lim}$, since $\Pi^{CONT}\to 0$ in the limit $k^2\to0$, 
$D_{\mu\nu}$ becomes negative, suggesting that the model is
not unitary in this region.

\begin{figure}[htbp]
\vspace{3pt}
\centerline{
\setlength\epsfxsize{280pt}
\epsfbox{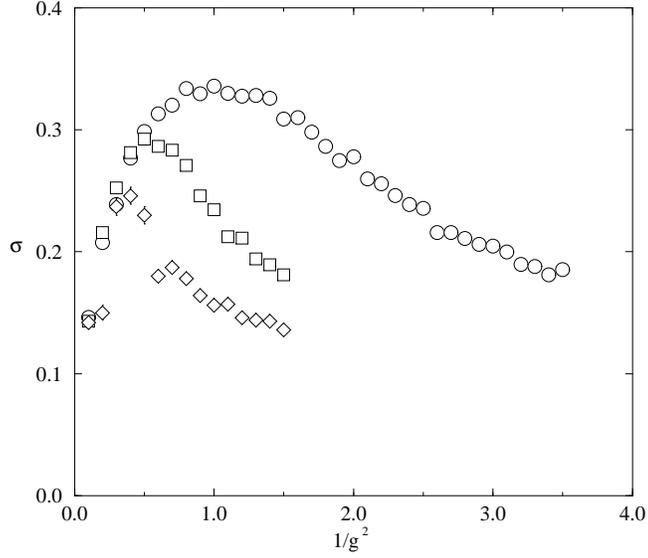}}
\caption{Chiral condensate $\sigma$ vs. $1/g^2$ for $N=1$
(circles), 2 (squares) and 3 (diamonds), showing discontinuous 
behaviour for $1/g^2\simeq0.3$.
\label{fig:hands_transition}}
\end{figure}

In Fig.~\ref{fig:hands_transition} we plot 
$\sigma=\langle\bar\chi\chi\rangle$ vs. $1/g^2$ for
$N=1,2,3$ and $m=0.10$ on a $8^3$ lattice~\cite{ldd96}. The three
models have apparently coincident condensates in the strong coupling
region $1/g^2\leq0.3$, which appear to be tending to zero in the 
$g^2\to\infty$ limit. It is tempting to associate this behaviour with
$g^2>g^2_{lim}$, but the correspondence with the value (\ref{eq:glim})
is not good. It may well be that the numerical value of the right-hand
diagram of Fig.~\ref{fig:diagrams} is considerably reduced in a chirally broken
vacuum; in any case we do not expect the $1/N$ expansion to be accurate
here for the reasons described above.

The ill-defined behaviour of the $g^2\to\infty$ limit of the
non-compact action (\ref{eq:non-compact}) presents no problem for most
of the
simulation results presented in this paper, which focus on the critical
region of the $N=1$ model around $1/g_c^2\simeq2.0$,
and of the $N=2$ model with $1/g_c^2\simeq0.6$. However, we expect
$g_c^2$ to increase as $N$ increases~\cite{Hong94,Itoh95,Kondo95,ldd96}, 
so that as
models with larger $N$ are explored,
eventually the relation between $g^2$ and $g_R^2$ will become 
important, particularly if lattice studies are to resolve the
issue of the existence and value of $N_{fc}$. It is therefore worth
making remarks on two other possible forms for the lattice Thirring
model.

For $N=1$ there is an alternative but equivalent way of introducing
the auxiliary field, which we will call the {\sl compact\/} form:
\bea
	S &=& \oh \sum_{x\mu i} \bar\chi_i(x) \eta_\mu(x) 
	\left(1+\sqrt{{2g^2\over N}}\exp(i\theta_\mu(x))\right)\chi_i(x+\hat\mu)
	+ \mbox{h.c.} \non \\
	  & & + m \sum_{xi} \bar\chi_i(x) \chi_i(x), 
\label{eq:compact}
\eea
where the link variables $\theta_\mu(x)$ are freely integrated 
over~\cite{Booth89,AliK95}. It is straightforward to perform this
integration 
to yield a term $\propto I_0(\sqrt{-2g^2\bar\chi(x)\chi(x+
\hat\mu)\bar\chi(x+\hat\mu)\chi(x)})$, with $I_0$ a modified Bessel
function. If we define $I_0$ by its power series expansion, then the
Grassmann nature of the $\chi,\bar\chi$ truncates the expansion
and forces equivalence with the partition function based on
(\ref{eq:lattice-thirring}). We have performed some trial simulations 
using (\ref{eq:compact}), and found identical results to those of
the non-compact form (\ref{eq:non-compact}) within statistical accuracy.
It is interesting to note in passing that using a similar method one can
show the equivalence of (\ref{eq:lattice-thirring}) with the strong
gauge-coupling (i.e. $\beta=0$) limit of the gauge-fermion-scalar 
($\chi U\phi$) model recently studied in connection with non-trivial
continuum limits in 3 and 4 dimensions~\cite{Frick95}.

The bilinear action (\ref{eq:compact}) can be extended to $N>1$ simply by
introducing explicit flavor indices $i$ on $\chi,\bar\chi$. Now,
however, terms of the form \hbox{$(\bar\chi(x)\chi(x+\hat\mu)\bar\chi
(x+\hat\mu)\chi(x))^n$} with $n>1$ are not suppressed 
by the Grassmann nature of
$\chi$: for $N$ flavors of staggered fermion there are thus $n$-point
couplings in the action with $n\leq4N$. Since ultimately we wish to
understand and contrast the behaviour of the Thirring model for various
values of $N$, we prefer the non-compact form
(\ref{eq:non-compact}), where the interaction is of the same form for
all $N$. It may well prove to be the case that in the RG
sense the higher-point couplings are irrelevant, and hence
(\ref{eq:non-compact}) and (\ref{eq:compact}) have identical continuum
limits. Both variants deserve study.

Finally we mention a recent numerical study of a gauge-invariant
version of the three dimensional Thirring model by Kim and 
Kim~\cite{Kim96}; the action is
\bea
	S &=& \oh \sum_{x\mu i} \bar\chi_i(x) \eta_\mu(x) 
	\exp(i\theta_\mu(x))\chi_i(x+\hat\mu)
	+ \mbox{h.c.} \non \\
	  & & + m \sum_{xi} \bar\chi_i(x) \chi_i(x)
       -{N\over g^2}\sum_{x\mu}\cos(\phi(x+\hat\mu)-\phi(x)+
\theta_\mu(x)), 
\label{eq:Kim}
\eea
with $\phi$ a gauge-covariant scalar field. This action has a local
gauge invariance by construction, corresponding to the hidden local 
symmetry noted in Refs.~\cite{Itoh95,Kondo95}.
In unitary gauge $\phi=1$
this action resembles (\ref{eq:non-compact}) in the long-wavelength
limit, but now has a periodic dependence on the auxiliary field. 
The main result of \cite{Kim96} is that the $N_f=6$ model appears
qualitatively different from $N_f=2$, once again supporting 
a value of $N_{fc}<6$.
It is an open question whether the action (\ref{eq:Kim}) lies in the
same universality class as (\ref{eq:lattice-thirring}).

We simulated dynamical fermions using a hybrid Monte Carlo (HMC)
algorithm. In all
cases we used a random trajectory length drawn from a Poisson
distribution with mean 0.9, and adjusted the timestep 
to maintain an acceptance rate of 70\% or greater. Typically we
performed measurements on configurations separated by 
between 2 and 5 trajectories. 
Our estimated
autocorrelation times measured on sequences of
configurations varied from $\sim1$ to $\sim3$ near the critical
coupling; all our quoted errors take autocorrelations into account.
The convergence
criterion we used in the conjugate gradient matrix inversion
was that the modulus of the residual vector
was $10^{-6}$ per lattice site during guidance and $10^{-9}$ per site 
for Metropolis and measurement. Unlike the case 
of the GN model~\cite{Hands93}, we were unable to boost 
acceptance rates by tuning the parameters of the guidance 
Hamiltonian. However, the vectorlike
nature of the interaction allows the simulation of the model for any value
of the number of lattice species $N$, corresponding to a number of physical 
fermions $N_f=2N$, via an even-odd partitioning. 
For the very same reason, the only diagonal term in the fermion matrix
comes from the mass term in the action. As $m$ gets small, we find that
the fermion matrix becomes very difficult to invert. Simulations with
small $m$ are therefore extremely time-consuming and represent a major
problem for the extrapolation of the data to the chiral limit.
Most of the results reported in this paper use between 200 and 500 
measurements at each value of the coupling and bare mass.
We have used periodic boundary conditions for the 
auxiliary field,  and for the fermions used periodic in the spacelike
directions and antiperiodic in the timelike direction. Most of the new
results in the paper are obtained on $16^3$ lattices; however we also
did some runs on an asymmetric $8^2\times16$ lattice, to give some 
estimate of finite volume effects on correlation function measurements.

In our simulations we measured the following observables:

\noindent{\it(i)\/} The chiral condensate
\bea
\sigma=\langle\bar\chi\chi\rangle={1\over V}\langle\mbox{tr}M^{-1}\rangle,
\eea
where $M$ is the fermion kinetic operator introduced in
eq.(\ref{eq:non-compact}). In Sec. 3 we will discuss numerical fits to
the model's equation of state using these measurements. To
enable alternative fitting procedures we list all our measurements made
on symmetrical (i.e. $L^3$) lattices for $N_f=2$ in Tabs. \ref{tab:Nf=2}
and \ref{tab:Nf=2_bis}, for $N_f=4$ in Tab. \ref{tab:Nf=4} and for
$N_f=6$ in Tab. \ref{tab:Nf=6}.

\noindent{\it(ii)\/} The longitudinal susceptibility
\bea
\chi_l &=& {1\over
V}\sum_x\langle\bar\chi\chi(0)\;\bar\chi\chi(x)\rangle_c \non \\
&=& {1\over V}\sum_x\bigl[\langle(\mbox{tr}M^{-1})(\mbox{
tr}M^{-1})\rangle
-\langle\mbox{tr}M^{-1}\rangle^2\bigr] 
-\langle\mbox{tr}(M^{-1}M^{-1})\rangle 
\label{eq:chil} \\
&\equiv& \chi_{ls}+\chi_{lns},\non
\eea
where we have denoted the components with disconnected and connected 
fermion lines respectively as ``singlet'' $\chi_{ls}$ and
``non-singlet'' $\chi_{lns}$ in analogy with QCD flavor dynamics.

\noindent{\it(iii)\/} The transverse susceptibility
\bea
\chi_t &=& {1\over
V}\sum_x\langle\bar\chi\varepsilon\chi(0)\;\bar\chi
\varepsilon\chi(x)\rangle_c \non \\
&=& \chi_{ts}+\chi_{tns} 
\label{eq:chit} \\
&\equiv& {1\over m}\langle\bar\chi\chi\rangle,\non
\eea
where we have used the fact that $\langle\bar\chi\varepsilon\chi\rangle$
vanishes, and the last line follows from a Ward identity.

\noindent{\it(iv)\/} The ``vector susceptibility''
\bea
\chi_v&=&{1\over V}\sum_{x,\mu=1,2}\langle
V_\mu(0)\;V_\mu(x)\rangle_c
\label{eq:vecsusc} \\
&=& \chi_{vs}+\chi_{vns},\non
\eea
where $V_\mu$ is the interaction current:
\bea
V_\mu(x)={i\over2}\bigl(\bar\chi(x)\eta_\mu(x)\chi(x+\hat\mu)+
\bar\chi(x+\hat\mu)\eta_\mu(x)\chi(x)\bigr).
\eea
Due to the inhomogeneous boundary conditions, we only averaged this quantity
over the spacelike directions.

All susceptibilities, of course, are identical to the bound state
propagator in the appropriate channel evaluated at zero momentum. The
non-singlet components were therefore easily evaluated from the bound
state correlators discussed below. The singlet parts, involving traces
over a single power of $M^{-1}$, were evaluated using gaussian noise
vectors. Since the susceptibility signals are intrinsically noisy, we
used
5 noise vectors per configuration, thus obtaining 5 independent
estimates of $\langle\bar\chi\chi\rangle$, and $5\times4/2=10$
independent susceptibility estimates.

In addition to these operators we have also performed spectroscopy
by evaluating timeslice 
propagators. We examined the following channels:

\noindent{\it(v)\/} Fermion propagator
\bea
C_f(t)=\mbox{Re}\sum_{x,y\;even}\langle\chi(0)\;\bar\chi(x,y,t)\rangle,
\label{eq:Cferm}
\eea
where note that the signal is improved if only sites an even number of
translations in the transverse directions are included in the sum.

\noindent{\it(vi)\/} Local composite operators
\bea
C_l(t)=\sum_{x,y}\langle\bar\chi\Gamma\chi(0)\;\bar\chi\Gamma\chi(x,y,t)
\rangle, 
\eea
where the space-dependent
phase $\Gamma$ is chosen to project onto various channels as
follows:
\bea
\mbox{``pion"}\;\;\;\Gamma(x)&=&\varepsilon(x),\label{eq:Cps}\\
\mbox{``scalar"}\;\;\;\Gamma(x)&=&1,\label{eq:Cs}\\
\mbox{``local vector"}\;\;\;\Gamma(x)&=&(-1)^{x+t}+(-1)^{y+t}.
\label{eq:Cvl}
\eea
The local vector operator is chosen by analogy with the rho-meson
operator commonly used in QCD simulations.

\noindent{\it(vii)\/} One-link composite operators
\bea
\mbox{``conserved vector"}\;\;\;C_{cv}(t)={1\over2}\sum_{x,y}
\sum_{\mu=1,2}\langle V_\mu(0)\;V_\mu(x,y,t)\rangle,
\label{eq:Cvc}
\eea
i.e. the two-point correlator between interaction currents $V_\mu(x)$,
corresponding to the non-singlet component of the vector susceptibility
(\ref{eq:vecsusc}). 

Of course, before assigning names to the various channels we should
examine the spin/flavor structure of the fermion bilinears using the
transformation (\ref{eq:b+b}). Since the staggered fermion action is
invariant only under translations by an even number of lattice
spacings, in general,  as $t\to\infty$, two-point correlators behave as
\bea
C(t)= Ae^{-M_d t}+B(-1)^te^{-M_at},
\label{eq:altfit}
\eea
i.e. displaying signals in both ``direct'' and ``alternating'' channels,
characterised by distinct masses $M_d$ and $M_a$ respectively. 
The spin/flavor assignments we find are shown in Tab.
\ref{tab:spinflav}. It is interesting to note that in three Euclidean
dimensions the local rho meson operator, which we have referred to as
``local vector'', does not in fact have vector quantum numbers.

In the calculation of timeslice propagators we evaluated the inverse
matrix $M^{-1}$ using a single local source on 
each configuration, chosen at a site randomly
displaced from the origin by an even number of lattice spacings in each
direction.

\section{The equation of state from RG equations}

Spontaneous breaking of chiral symmetry is signalled by a non-vanishing
chiral condensate as $m\rightarrow 0$. As already pointed out in 
the previous section, it is extremely difficult to run the HMC for very
small values of the bare mass. Therefore we are bound to work with 
a restricted set of values for $m$, making an extrapolation to $m=0$ 
at a fixed value of $g$ rather unreliable.

In order to find numerical evidence of the RG structure described
in the introduction, the scaling properties of the theory at the UV fixed 
point are going to be used to find an equation of state allowing a
global
fit of the data. The equation of state relates the order parameter for
the broken symmetry (in our case the chiral condensate) to the value of
the external symmetry breaking field (the bare mass $m$) as a function
of the external parameters (coupling constant, finite size of the lattice),
thus providing a powerful tool for investigating the behaviour of the order
parameter near the phase transition in the $(g,m)$--plane.

\subsection{Scaling properties at fixed lattice size}

Let us briefly recall how an equation of state can be obtained from the RG
equations (RGE) (see eg. \cite{ZJ}). Consider a theory of a bosonic 
field $\phi$ regularised with 
some cut--off $\Lambda$, coupled to an external field $H$ and let 
$M=\langle\phi\rangle$ be the order parameter. Then the effective 
potential can be expanded in 
powers of $M$:
\beq
	\Gamma(M,t,\Lambda) = \sum \frac{M^n}{n!} \Gamma^{(n)}
	(p_i=0,t,\Lambda)
\eeq
where $\Gamma^{(n)}$ are the 1PI Green functions at zero external momenta,
$t$ is the reduced coupling (see (\ref{eq:redt}))
and $\Lambda$ the cut--off.
The external field, for given magnetisation, can be written as:
\beq 
	H = \frac{\partial\Gamma}{\partial M} = 
    	\sum \frac{M^n}{n!} \Gamma^{(n+1)}(p_i=0,t,\Lambda)
\eeq
The above equation allows the RGE for $H$ to be written:
\beq
	\left\{ \Lambda \de{\Lambda} + \beta(t) 
	\de{t} - \oh \eta(t) \left( 1 + M\de{M} \right) \right\}
	H(t,m,\Lambda) = 0
\eeq
where:
\bea
	\beta(t) &=& \Lambda \de\Lambda g \\
        \eta(t)  &=& - \Lambda \de\Lambda \log Z
\eea
and $Z$ is the wave-function renormalisation.

The solution of the RGE is obtained by using the method of characteristics. 
In a neighbourhood of the fixed point, the solution can be combined with 
dimensional analysis in order to rescale the cut-off and obtain the 
general result:
\beq
	H(M,t,1) \sim M^\delta {\cal F}(t M^{-1/\beta})
\label{eq:scaling1}
\eeq
where $\cal F$ is a universal scaling function. 
By setting $H=0$ in eq.~(\ref{eq:scaling1}), the critical behaviour of the 
order parameter when the external field is switched off is recovered:
\beq
	t M^{-1/\beta} \sim \mbox{const}
\eeq
while, for $t=0$, ${\cal F}(0)$ is a constant and hence:
\beq
	H \sim M^{\delta}
\eeq
showing clearly that $\beta$ and $\delta$ are the usual critical exponents
introduced in the context of phase transitions.
If the critical exponents  are related to the existence of a UV fixed point, 
as we are assuming in this section, they must obey the hyper-scaling 
relations, obtained under the assumption that there is a single physical
length scale near the continuum limit:
\bea
	\beta  &=& \oh \nu (d-2+\eta) \non \\
	\delta &=& \frac{d+2-\eta}{d-2+\eta} \label{eq:beta} 
\eea
where $\eta$ is the anomalous dimension and $\nu$ is the critical exponent
which characterises the divergence of the correlation length as 
$t\rightarrow 0$.

For the Thirring model the reduced coupling $t$ is identified with:
\beq
	t = 1/g^2 - 1/g_c^2
\label{eq:redt}
\eeq
and the symmetry breaking field with the bare mass $m$, while the order
parameter is the chiral condensate 
$\langle\bar\chi\chi\rangle$ defined in the previous section.
A Taylor expansion for small $t$ reduces Eq.~(\ref{eq:scaling1}) to our 
first equation of state:
\beq
	m = B \cond^\delta + A t \cond^{\delta-1/\beta} + 
	{\cal O}\left((t\cond^{-1/\beta})^2\right)
\label{eq:eos1}
\eeq

If the critical behaviour is described by mean-field theory, the critical 
exponents in Eq.~(\ref{eq:scaling1}) are expected to be $\delta=3$ and
$\beta=1/2$, leading to:
\beq
\cond^2 = \frac{1}{B} \frac{m}{\cond} - \frac{At}{B}
\eeq
In mean-field theory the square of the chiral condensate is a linear 
function of the ratio $m/\cond$. A plot of $\cond^2$ vs. $m/\cond$ is
called a Fisher plot. In such a plot a positive value of the intercept 
corresponds to a non-vanishing condensate for $m=0$, i.e. to chiral
symmetry breaking, while the value of the intercept will be exactly zero 
at the critical coupling. 

Eq.~(\ref{eq:eos1}) permits a five parameter fit of
the numerical data for the chiral condensate.
In order to reduce the number of free parameters in the fit, one can
set $\delta-1/\beta=1$ and keep $\delta$ as the only exponent to be
determined from the data. This relation originally arose 
from solution of the
SD equations for the gauged Nambu -- Jona-Lasinio model in four
dimensions in ladder approximation \cite{Kocic91}; here we simply use
it as a plausible hypothesis to be tested against the data.
In what follows we refer to the 
5 and 4 parameter fits as fits I and II respectively. 
The outcome of fit I shows that the above relation between $\delta$
and $\beta$ is satisfied wihin errors. 

Since both formul\ae ~rely on a Taylor expansion around $t=0$, they are
only expected to describe the behaviour of the chiral condensate in 
a neighbourhood of the critical coupling, so that only a reduced set
of values of $g$ can be fitted by our equation of state. The number  
of points included in the fit has been chosen in such a way to 
minimise the $\chi^2$ value.

\begin{figure}[htbp]
\vspace{3pt}
\psdraft
\centerline{
\setlength\epsfxsize{300pt}
\epsfbox{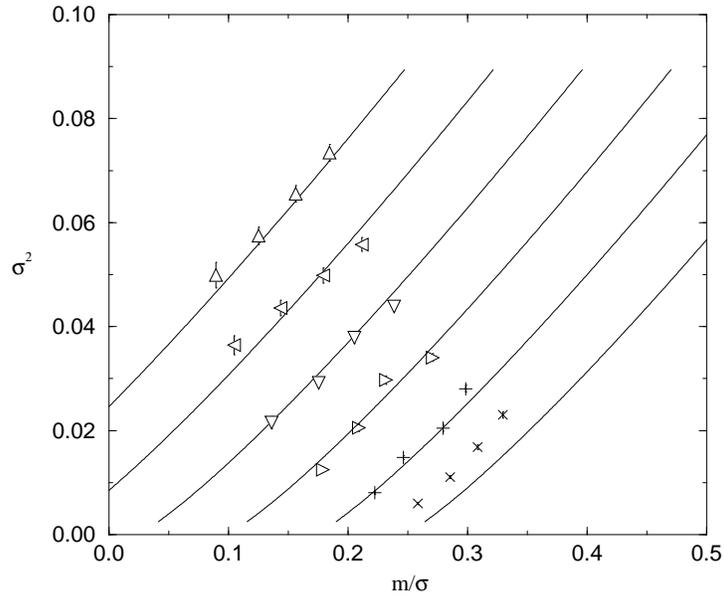
}}
\psfull
\caption{Fisher plot for $N_f=2$ on $12^3$ lattice. 
Values of $1/g^2$ range from
1.6 ($\triangle$) in steps of 0.2 to 2.6 ($\times$).
\label{fig:fish12.2}}
\end{figure}

\begin{figure}[htbp]
\vspace{3pt}
\psdraft
\centerline{
\setlength\epsfxsize{300pt}
\epsfbox{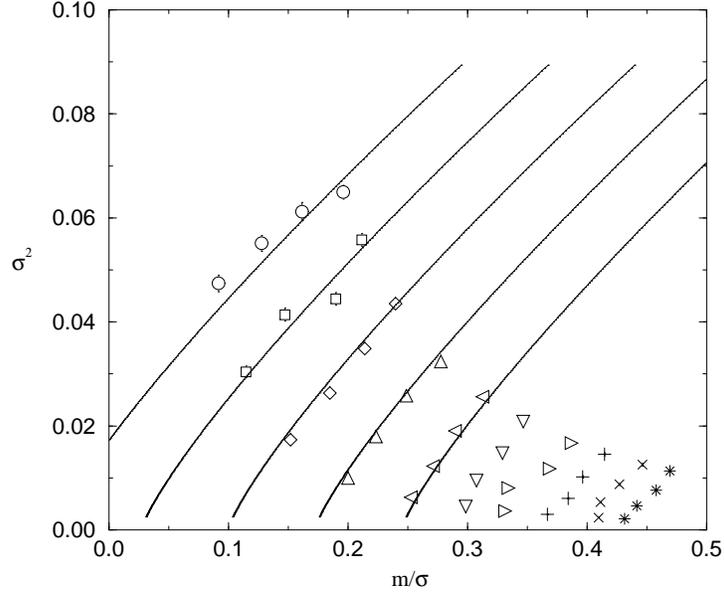
}}
\psfull
\caption{Fisher plot for $N_f=4$ on $12^3$ lattice. Values of $1/g^2$ range from
0.5 ($\bigcirc$) in steps of 0.1 to 1.4 ($\ast$).
\label{fig:fish12.4}}
\end{figure}

\begin{figure}[htbp]
\vspace{3pt}
\psdraft
\centerline{
\setlength\epsfxsize{300pt}
\epsfbox{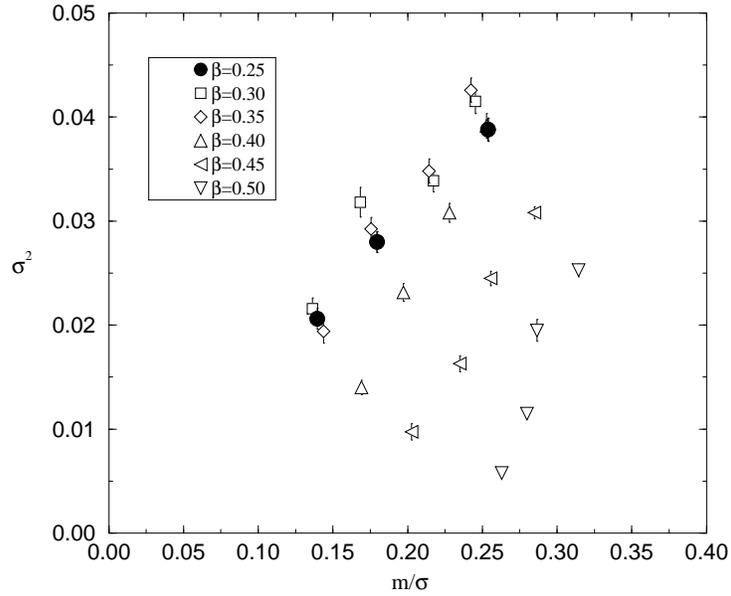
}}
\psfull
\caption{Fisher plot for $N_f=6$ on $12^3$ lattice.
\label{fig:Fisher_6}}
\end{figure}

In Figs.~\ref{fig:fish12.2} and \ref{fig:fish12.4} 
we show the Fisher plots for $N_f=2,4$ on the $12^3$ lattice,
together with the curves obtained using
the parameters of fit II. The existence of a critical point separating a 
chirally symmetric phase from a phase where chiral symmetry is spontaneously
broken can be easily spotted from the figures according to the above
criterion. The corresponding lattice data can be fitted to the equation
of state, yielding stable results and reasonably small $\chi^2$. 
The detailed results from the fits are
reported in Tabs.~\ref{tab:fit1} and~\ref{tab:fit2}. 
Fits I and II agree within errors for both values of $N_f$ on
the $12^3$ lattice, justifying the use of the fit with the reduced number
of parameters. From a practical point of view, such a justification
becomes very important since a smaller number of variables is more easily
fitted from the relatively small datasets that one 
can produce when dealing with dynamical fermions. In the following the 
results quoted will always be the ones from fit II, unless specified.
The critical coupling $g_c$ does not seem to be unduly affected by finite-size
effects, as one can see by comparing the results in Tab.~\ref{tab:fit1},
\ref{tab:fit2}.
The values extracted from the $12^3$ data for $N_f=2,4$ can thus
be used to check the SD prediction of equation 
(\ref{eq:HongPark}):
\beq
\frac{g^2_c(N_f=2)}{g^2_c(N_f=4)} = 0.291
\eeq
Comparing with our result
\beq
\frac{g^2_c(N_f=2)}{g^2_c(N_f=4)} = 0.342 \pm 0.015
\eeq
we can conclude that the above prediction is more than 
three standard deviations away from the lattice data.
This seems to favour the scenario where a critical number of flavors 
$N_{fc}$,  exists.
For  $N_f>N_{fc}$ 
the theory would then be chirally symmetric for any value of the coupling.
This conjecture is also supported by our data for $N_f=6$, shown in 
Fig.~\ref{fig:Fisher_6}, where it is very difficult to identify 
a critical point corresponding to a symmetry breaking phase transition.
Qualitatively as $g$ increases the trajectories seem to be accumulating 
around a line which if continued would intercept the horizontal axis.
Quantitatively, neither fit I nor fit II provides a satisfactory 
description of the data.
It is interesting
to note that, in all the cases where the fit converged, the critical 
exponent $\delta$ is at least
two standard deviations away 
from the mean field value $\delta=3$.
Finally, let us remark that finite-size effects influence the value of
the critical exponent $\delta$: the next sub--section is dedicated to the
inclusion of these effects in our equation of state, so that the 
exponents extracted by our fitting procedure are those corresponding to
the thermodynamic limit.

\subsection{Finite size scaling and the equation of state}

Finite size effects can be used to extract further information
about the nature of the critical point.
In the usual approach to 
finite size scaling, the RGE are assumed to be completely 
insensitive to finite
size effects, since renormalisation deals with the short distance 
properties of the theory; they are therefore unchanged in a
finite geometry. Nonetheless the finiteness of the volume 
affects the solutions of RGE, which now depend on one more dimensionful
parameter. This is equivalent to considering the inverse linear size of the 
lattice, $1/L$, as an additional 
relevant scaling field with eigenvalue $1$ close to a fixed point
at $L\rightarrow\infty$. 
By repeating the arguments explained above, we obtain an equation 
expressing the external field
as a universal function of two rescaled variables:
\beq
	m \cond^{-\delta} \sim  {\cal F}(t M^{-1/\beta}, 
	L^{-1/\nu} \cond^{-1/\beta})
\label{eq:scaling2}
\eeq
which can again be expanded to yield a new equation of state,
\beq
	m = B \cond^\delta + A (t + C L^{-1/\nu}) \cond^{\delta-1/\beta} 
	+ \mbox{higher order terms},
\label{eq:eos2}
\eeq
to which the data can be fitted, in principle
allowing the extraction of three critical exponents
($\delta$, $\beta$ and $\nu$) and the critical coupling, taking into 
account finite size effects. The parameters obtained from this 
fit are the thermodynamic (infinite volume) limit ones.
Hence, assuming the existence of a UV fixed point, the fit to this new
equation of state provides a method to determine the thermodynamic limit
of the critical exponents and coupling without further
hypothesis. 

Once again, it is important to stress that Eq.~(\ref{eq:eos2}) is based
on a Taylor expansion and is expected to describe the behaviour of
the data only in a restricted region around the fixed point. As in the
previous section, the set of data included in the fit is selected by
maximising the $\chi^2$/d.o.f. In order to reduce the number of
parameters in the fit, $\delta-1/\beta$ is set to 1 again and
$\nu$ is assumed to obey the hyperscaling relation~(\ref{eq:beta}),
as it should be in the vicinity of a RG fixed point
so that $\nu=(\delta+1)/3(\delta-1)$. The 
number of parameters is then reduced from seven to five.
Data from lattice sizes ranging from 8 to 16 have been included in
the fit, whose results are reported in Tab.~\ref{tab:fit3}. It is 
clear from the values of the critical exponents that the theory is 
not mean-field and that the data around the critical point seem
to match well the expected behaviour of a UV fixed point.

\begin{figure}[htb]
\psdraft
\centerline{
\setlength\epsfxsize{300pt}
\epsfbox{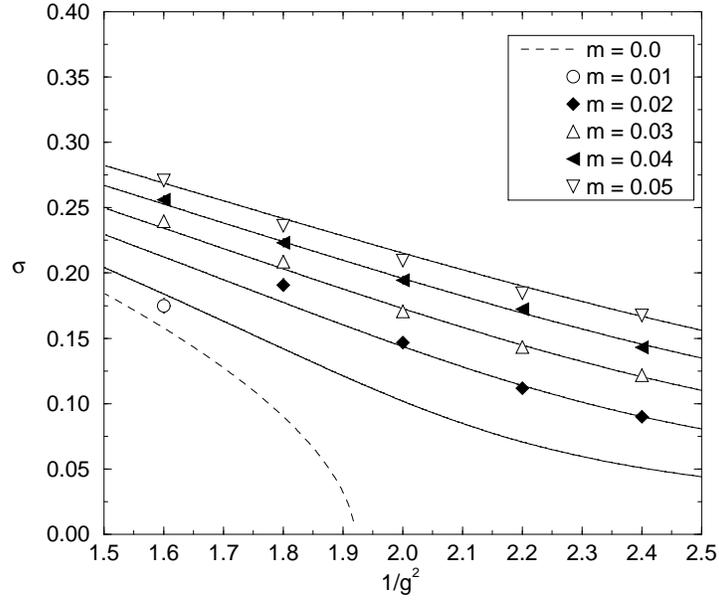
}}
\psfull
\caption{Chiral condensate vs. $1/g^2$ on $12^3$ lattice. 
The solid lines are the fits from (\ref{eq:eos2}), the dashed line 
the extrapolation to chiral and thermodynamic limits.
\label{fig:chi12}}
\end{figure}

\begin{figure}[htb]
\psdraft
\centerline{
\setlength\epsfxsize{300pt}
\epsfbox{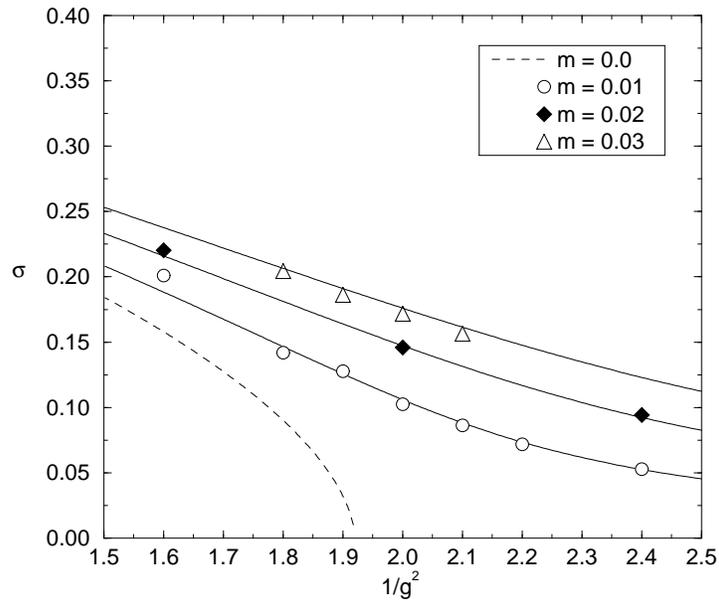}}
\psfull
\caption{Chiral condensate vs. $1/g^2$ on $16^3$ lattice.
\label{fig:chi16}}
\end{figure}

The data for the chiral condensate vs. $1/g^2$ are reported in 
Fig.~\ref{fig:chi12} and~\ref{fig:chi16} for the $12^3$ and $16^3$
lattices respectively, together with the curves obtained from the fit.
The dashed line represents the chiral condensate for $m=0$ in the 
thermodynamic limit. 

To conclude, in this section we have shown that the chiral condensate
data of the $N_f=2$ and $N_f=4$ models support  the 
hypothesis of a power-law equation of state 
in the critical region. The fits obtained were also 
consistent with the constraint $\delta-1/\beta=1$.
In passing it is worth remarking that it is not possible to obtain a 
good fit to our data with a mean-field equation of state with 
logarithmic corrections, as used for ${\rm QED}_4$ in \cite{Gock96}. 
The critical exponents extracted for the two models 
are distinct, in neither case resembling those of mean field theory,
suggesting that the two models have continuum limits described by 
distinct field theories. Of
course, this in itself is not direct 
evidence for an interacting continuum limit
at the fixed point; 
to establish that would require the study of lines of constant 
physics in the space of bare parameters $g$ and $m$, and measurement of 
$n$-point functions in the critical region. No
similar fit was found for $N_f=6$, suggesting that in this case there is
no transition, and that $N_{fc}<6$ for the Thirring model.

\section{Susceptibilities and Spectroscopy}

In this section we review results of susceptibility and spectroscopy
measurements, principally on the $N_f=2$ model studied on a $16^3$
lattice with $m=0.01$ (ie. unless stated otherwise), 
but also showing some results from a smaller lattice, 
and for $N_f=4$, for comparison. First we consider 
susceptibilities, generically
denoted $\chi$, which as discussed in Sec. 2 are equivalent to bound
state propagators evaluated at zero three-momentum. They take
large values whenever there are long-range correlations present.
There are two reasons
why $\chi$ is a useful quantity to study. Firstly, it can be separated
into singlet $\chi_s$ and non-singlet $\chi_{ns}$ components; $\chi_s$,
formed from disconnected valence fermion lines, is thus a useful
measure of the importance of sea fermion loops in the vacuum. Since
there are no dynamical bosons in the original model, any long-range
interactions must be mediated by fermion ``bubbles''. The relative
strength of $\chi_s$ versus $\chi_{ns}$ is the simplest probe of this
effect.  Secondly, if we assume that fermion -- anti-fermion composites
resemble canonical bosons, then the relation
\bea
\chi={Z\over M^2}
\label{eq:susc}
\eea
holds, where $M$ is the bound state mass and $Z$ is a renormalisation
constant associated with the projection of the composite operator onto
the propagating state. Hence a large susceptibility signals a small
bound state mass. However, in a non-confining theory the assumptions
leading to (\ref{eq:susc}) may not be justified, eg. when there
is a contribution from the fermion -- anti-fermion continuum as in the scalar
channel of the three dimensional GN model~\cite{Hands93}. In
this case extraction of the bound-state mass is model dependent, whereas
the susceptibility is still unambiguously defined.

\begin{figure}[htbp]
\vspace{3pt}
\centerline{
\setlength\epsfxsize{300pt}
\epsfbox{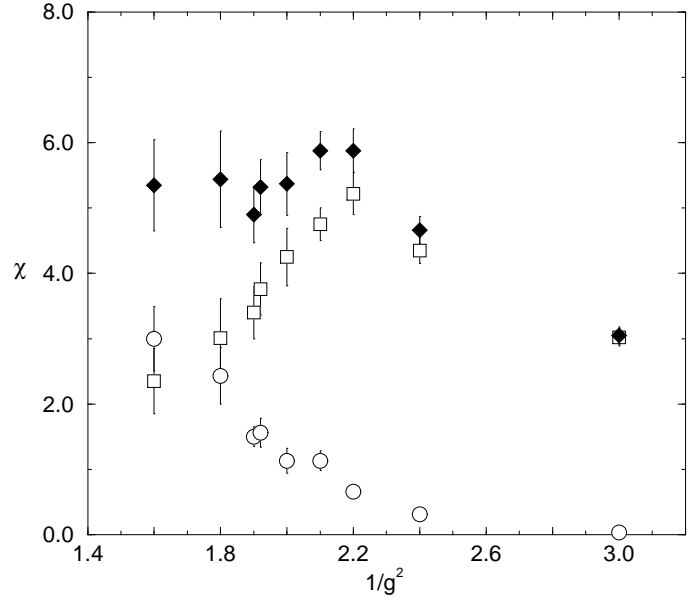}}
\caption{$\chi_{ls}$ (circles),
$\chi_{nls}$ (squares) and $\chi_{l}$ (diamonds) vs. $1/g^2$.
\label{fig:chil}}
\end{figure}

First we consider the scalar channel,
plotting in Fig.~\ref{fig:chil} $\chi_{ls}$, $\chi_{lns}$, and $\chi_l$ versus
$1/g^2$ (see Eq.~(\ref{eq:chil})). In the symmetric phase the
signal is dominated by $\chi_{lns}$, which peaks at
$1/g^2\sim2\;-\;2.2$, indicating a possible light scalar state, before
falling away in the broken phase. The singlet contribution is almost
absent in the symmetric phase, but rises across the transition to
dominate the signal by $1/g^2=1.6$. The total longitudinal
susceptibility is approximately constant in the broken phase.

\begin{figure}[htbp]
\vspace{3pt}
\centerline{
\setlength\epsfxsize{300pt}
\epsfbox{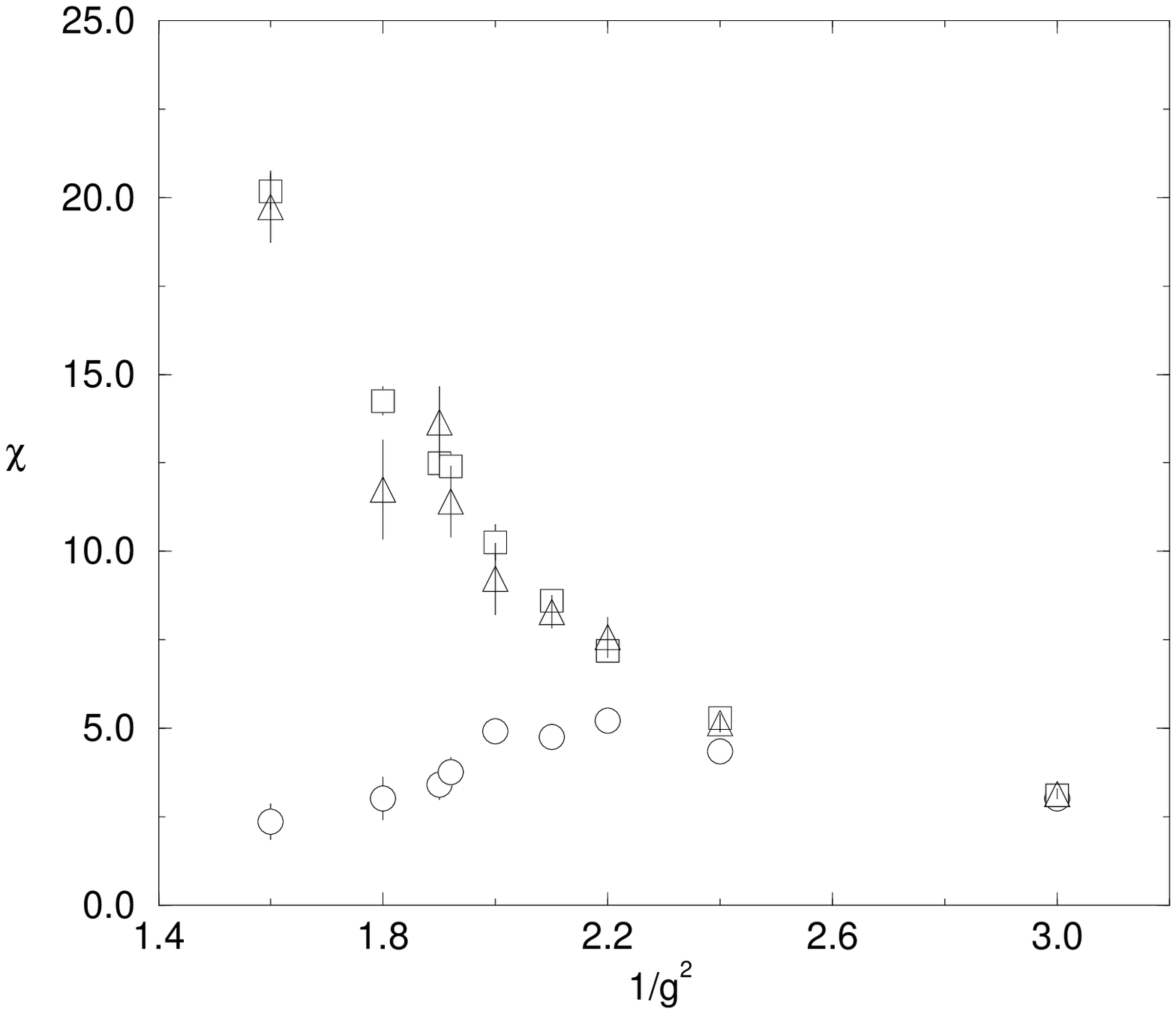}}
\caption{
$\chi_t$ (squares), $\chi_{tns}$ (triangles) and $\chi_{lns}$
(circles) vs. $1/g^2$ for $N_f=2$.
\label{fig:chilt2}}
\end{figure}

\begin{figure}[htbp]
\vspace{3pt}
\centerline{
\setlength\epsfxsize{300pt}
\epsfbox{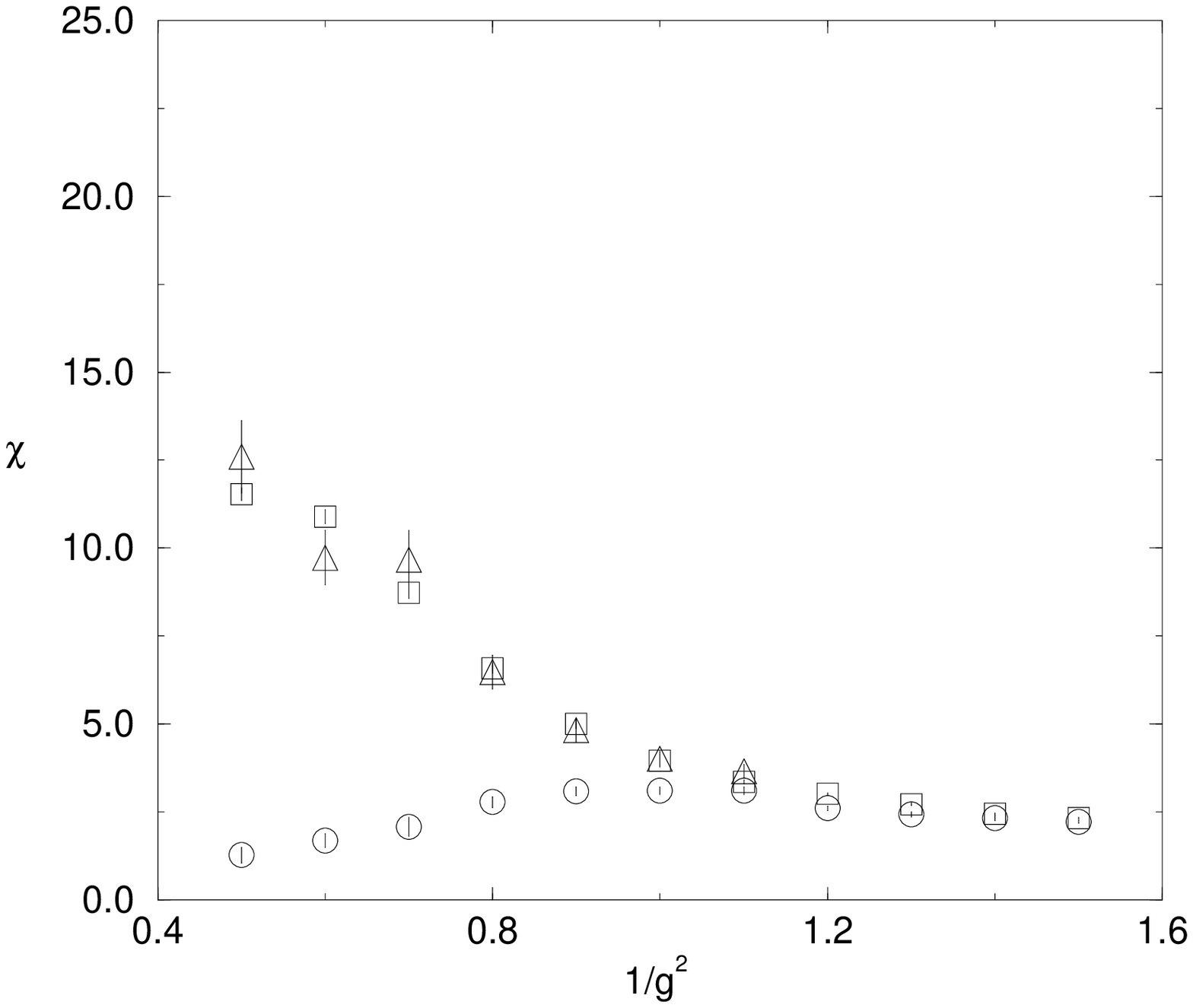}}
\caption{
$\chi_t$ (squares), $\chi_{tns}$ (triangles) and $\chi_{lns}$
(circles) vs. $1/g^2$ for $N_f=4$.
\label{fig:chilt4}}
\end{figure}

Next we compare longitudinal and transverse susceptibilities
(\ref{eq:chil},\ref{eq:chit}), showing data for $N_f=2$, $m=0.01$ on a
$16^3$ lattice in Fig.\ref{fig:chilt2} and for $N_f=4$, $m=0.02$ 
on a $12^3$ lattice
in Fig.~\ref{fig:chilt4}. 
The main features are that $\chi_l\simeq\chi_t$ deep in the
symmetric phase, but that $\chi_t$ rises steeply in the broken phase
corresponding to the appearance of a massless Goldstone pion. Note that
the non-singlet contribution $\chi_{tns}$ saturates the full Ward
identity result within errors; this is as expected since singlet --
non-singlet degeneracy could only be lifted by an axial anomaly, which
is absent both for staggered fermions and for three dimensions.

\begin{figure}[htbp]
\vspace{3pt}
\centerline{
\setlength\epsfxsize{300pt}
\epsfbox{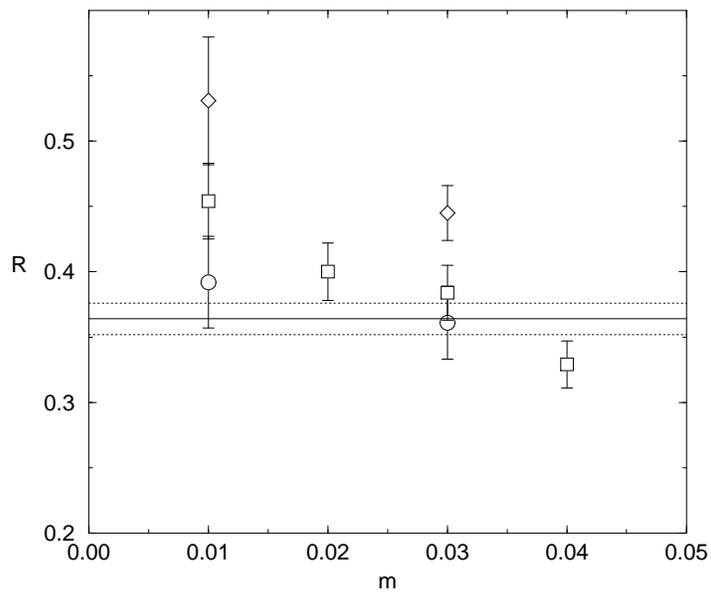}}
\caption{Susceptibility ratio $R$ vs. $m$ for $1/g^2=1.9$ (circles),
1.92 (squares) and 2.0 (diamonds). The horizontal band marks the value
of $1/\delta$ from the fit (\ref{eq:eos2}).
\label{fig:chi_ratio}}
\end{figure}

There is an interesting relation between the ratio $R$ of longitudinal
to transverse susceptibilities and the critical scaling behaviour of the
model, first noted in~\cite{KKW93} and subsequently developed 
in~\cite{Gock96}. Using the action (\ref{eq:lattice-thirring}) and the 
Ward identity (\ref{eq:chit}), we have
\bea
R(m,g^2)={\chi_l\over\chi_t}={m\over{\langle\bar\chi\chi\rangle}}
{{\partial\langle\bar\chi\chi\rangle}\over{\partial m}}.
\eea
Therefore $R$ can be calculated using the 
proposed equation of state (\ref{eq:eos1}):
\bea
R(m,g^2)={1\over{\left(\delta-{\displaystyle1\over\displaystyle\beta}\right)+
{\displaystyle{B\langle\bar\chi\chi\rangle^\delta}\over
\displaystyle{\beta m}}}}.
\eea
In general $R$ depends on $m$, but its limiting forms in the chiral
limit are easily evaluated.
In the broken phase, $\langle\bar\chi\chi\rangle$ remains non-vanishing
in the chiral limit, whereas in the symmetric phase
$\langle\bar\chi\chi\rangle^\delta$ vanishes faster than $m$; hence
\bea
\displaystyle\lim_{m\to0}R=\cases
{0,&$g^2>g_c^2$;\cr {1\over{\delta-{1\over\beta}}},&$g^2<g_c^2$.\cr}
\eea
The near-degeneracy of $\chi_l$ and $\chi_t$ in the symmetric
phase from $1/g^2=2.4$ onwards seen in Fig.~\ref{fig:chilt2}, 
and from $1/g^2=1.2$ onwards in Fig.~\ref{fig:chilt4},
therefore provides additional support 
for our preferred value of $\delta-1/\beta=1$ in the equation of state
(\ref{eq:eos1}).
Exactly at the critical coupling, however, $R$ becomes independent of
$m$, and we have
\bea
R\vert_{g=g_c}={1\over\delta}.
\eea
In Tab. \ref{tab:R} and Fig.~\ref{fig:chi_ratio} 
we plot $R$ for various values of $m$ and $g$ near the critical
point, in particular for four values of $m$ evaluated at our best
estimate of the critical coupling $1/g^2=1.92$. 
It is worth noting that the ratio
$\chi_{ls}/\chi_{lns}$ exceeds 40\% in this region, showing the
importance of including the disconnected contributions in this
particular calculation.
However, we also find 
that $R$ decreases significantly with $m$, at first
sight implying that the model is in the symmetric phase at this
coupling; the variation of $R$ with $m$ is less pronounced at
$1/g^2=1.90$, which would thus appear to be closer to the critical
coupling. This apparent contradiction is resolved if we take the
finite-size scaling form (\ref{eq:eos2}) seriously. Using the 
fit parameters from Tab. \ref{tab:fit3} the coefficient
multiplying the $\langle\bar\chi\chi\rangle^{\delta-1/\beta}$ term
actually vanishes for $1/g^2\simeq1.88$. We should therefore expect to
observe a constant $R$ at slightly stronger couplings than the critical
one on a finite system. Although we have insufficient data to verify
this directly, it is clear that this kind of 
analysis has the potential to provide an
independent estimate of $g_c^2$ and $\delta$.
An optimistic reader would
consider the near-agreement found here  
as a signal of the fact that the RG structure
underlying the equation of state approach is indeed realised in the 
Thirring model.

\begin{figure}[htbp]
\vspace{3pt}
\centerline{
\setlength\epsfxsize{300pt}
\epsfbox{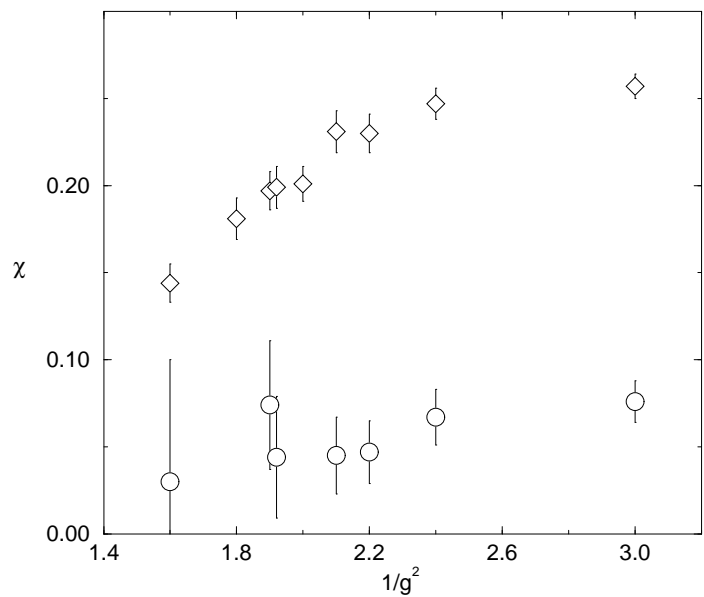}}
\caption{Vector susceptibilities $\chi_{vs}$ (circles) and $\chi_{vns}$
(diamonds) vs. $1/g^2$.
\label{fig:vec_susc}}
\end{figure}

In Fig.~\ref{fig:vec_susc} we plot the vector susceptibilities $\chi_{vs}$ and
$\chi_{vns}$ (\ref{eq:vecsusc}). The data is quite noisy; the trends are
that $\chi_{vns}$ appears roughly constant in the symmetric phase and
fall away in the broken phase, and that the ratio $\chi_{vs}/\chi_{vns}$
lies between 0.2 and 0.3 throughout the region.

Next we discuss spectroscopy, which involves the calculation of the 
timesliced
correlators $C(t)$ defined in Sec.~2.
We follow the usual procedure of extracting the mass of the 
lowest--lying state contributing to the two--point functions from 
their single exponential decay for large time separations.
In performing our minimum $\chi^2$ fits, we include correlations between
the data at different timeslices. We used the bootstrap technique to
estimate errors on the fitted parameters. We present results from 
$8^2\times 16$ and $16^3$ lattices.

First consider the fermion, pion 
and scalar channels. The fermion, since it cannot mix with
any excited states, shows the clearest signal and
can be fitted over the entire range by the form
\bea
C_f(t)=A\left(e^{-\mu_Rt}-(-1)^t e^{-\mu_R(L-t)}\right),
\eea
where $\mu_R$ is the physical fermion mass and
the minus sign between the forward and backward terms is due to our 
choice of antiperiodic boundary conditions in the timelike direction. 
All the quoted results
are obtained from fitting in the range $t\in[2,13]$ with a typical 
$\chi^2$/dof $\sim 1$. Varying the fit range produces negligible 
variations in the fitted parameters and $\chi^2$/dof.

\begin{figure}[htbp]
\vspace{3pt}
\centerline{
\setlength\epsfxsize{250pt}
\epsfbox{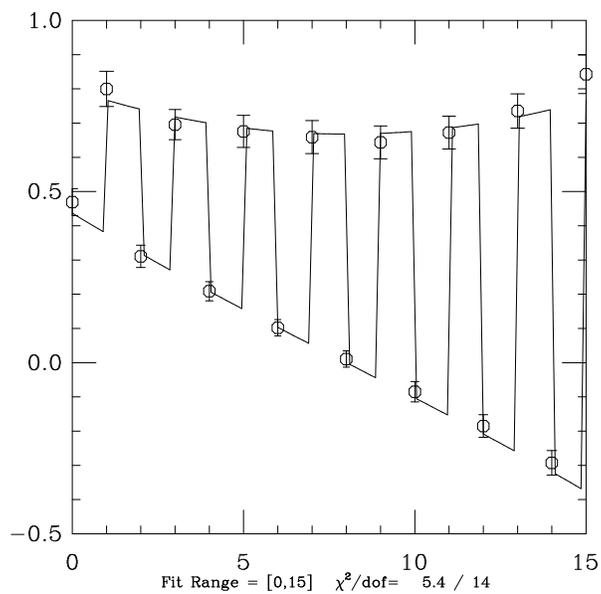}}
\caption{Fermion correlator for $1/g^2=2.2$.
\label{fig:ferm_corr}}
\end{figure}

\begin{figure}[htbp]
\vspace{3pt}
\centerline{
\setlength\epsfxsize{250pt}
\epsfbox{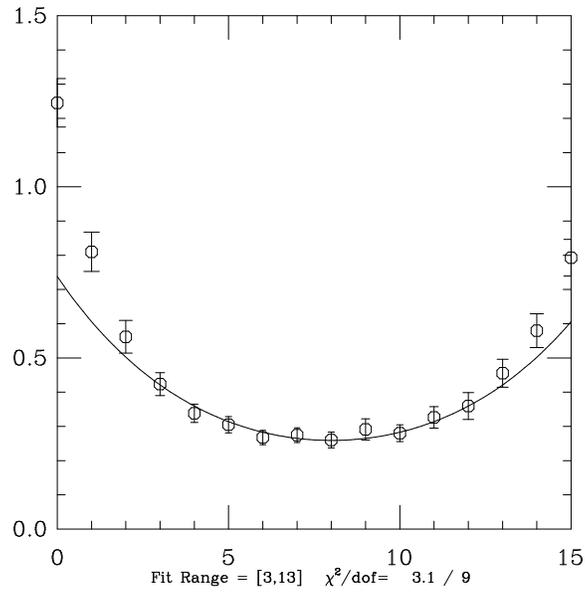}}
\caption{Pion correlator for $1/g^2=2.2$.
\label{fig:pion_corr}}
\end{figure}

\begin{figure}[htbp]
\vspace{3pt}
\centerline{
\setlength\epsfxsize{250pt}
\epsfbox{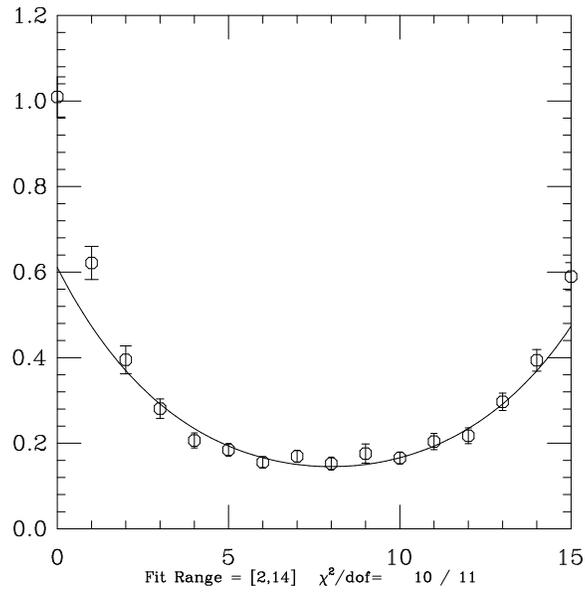}}
\caption{Scalar correlator for $1/g^2=2.2$.
\label{fig:sca_corr}}
\end{figure}

Both the scalar and pion channels were
fitted by the form
\bea
C_l(t)=A\left(e^{-Mt}+e^{-M(L-t)}\right).
\eea
The pion exhibits single exponential decay from early timeslices and
it is possible to fit the data in the range $t\in[3,13]$ in almost all
cases, again with a typical $\chi^2/\mbox{dof}\sim 1$. Similarly to the 
fermion case, fitting over a smaller time interval within this range
produces minimal variations. 
The scalar channel is more problematic. Firstly the signals exhibit greater
statistical noise. Secondly, as distinct from the pion channel, we can
only fit the correlators in a restricted time interval, suggesting 
contamination of the signal by higher-mass states. On the $8^2\times 16$
lattice it is not possible to reliably extract the ground-state signal
for $m=0.05$, while for $m=0.02$, fits are possible but quite sensitive to 
the fitted range. 
In this latter case we therefore choose a conservative fit range
$t\in [4,11]$; fitting outside this range leads to a much higher $\chi^2$/dof.
The fits on the $16^3$ data are more stable for all values of the fermion
bare mass. In the strong coupling regime, where the scalar mass gets 
larger, it proves harder to extract the ground--state mass regardless
of bare mass or lattice size, resulting in larger errors
on the fitted masses in this phase. In order to improve on this behaviour,
it would be necessary to move to larger lattices. With regard to this last 
point, our present results should therefore be considered as an attempt
to find qualitative evidence of the phase structure of the model, rather
than a precise evaluation of its spectrum.
The detailed results
are presented in Tabs.~\ref{tab:16_3_001_mass}, \ref{tab:16_3_002_mass},
\ref{tab:16_3_003_mass}, \ref{tab:8_2_16_002_mass} and 
\ref{tab:8_2_16_005_mass}. Examples of the fermion, pion and scalar
signals are shown in Figs.~\ref{fig:ferm_corr}, \ref{fig:pion_corr} and
\ref{fig:sca_corr} respectively.

A comparison between Tabs. \ref{tab:16_3_002_mass} and 
\ref{tab:8_2_16_002_mass} enables a crude assessment of finite volume effects.
In the smaller volume the fermion mass $\mu_R$ is shifted upwards 
significantly in the symmetric phase, while remaining relatively unaffected,
within errors, deep in the broken phase. The pion, in contrast, is 
lighter in the smaller volume, the effect again being more pronounced in the 
symmetric phase, but still significant in the broken phase. 
The scalar mass seems
the least affected by the finite volume.

\begin{figure}[htbp]
\vspace{3pt}
\centerline{
\setlength\epsfxsize{300pt}
\epsfbox{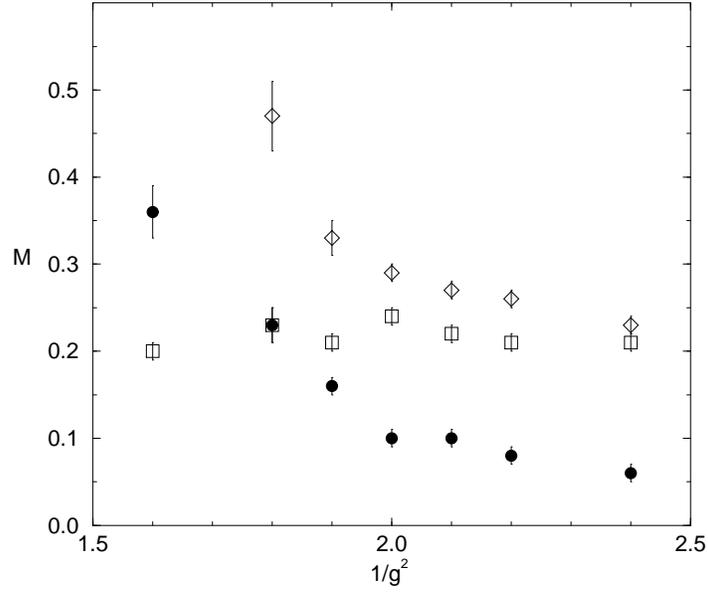}}
\caption{Fermion (filled circles), pion (squares) and scalar (diamonds)
masses vs. $1/g^2$.
\label{fig:masses}}
\end{figure}

\begin{figure}[htbp]
\vspace{3pt}
\centerline{
\setlength\epsfxsize{300pt}
\epsfbox{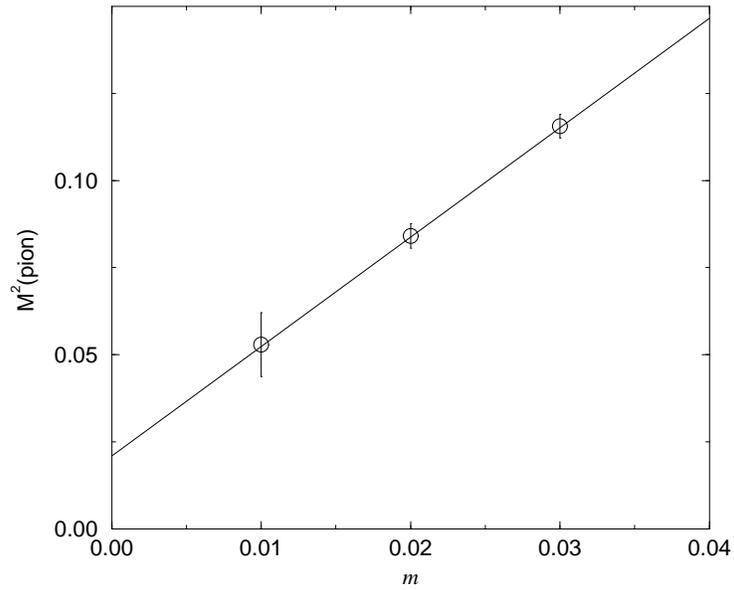}}
\caption{$M_\pi^2$ vs. $m$ in the broken phase at $1/g^2=1.8$.
\label{fig:pion_vs_bare}}
\end{figure}

In Fig. \ref{fig:masses} we plot the measured masses against $1/g^2$ for 
the $16^3$ system at $m=0.01$.
While the pion mass stays
constant, the fermion mass is small in the symmetric phase, but rises
steeply across the transition to become heavier than the pion in the
broken phase. Just on the symmetric side of the transition we have
$M_\pi\simeq2\mu_R$, indicating that the pion is a weakly bound state --
unfortunately by $1/g^2=2.4$ this picture breaks down, perhaps due to
finite volume effects. The scalar mass, meanwhile, is just greater but of the 
same order as the pion mass in the symmetric phase, 
but rises across the transition to become appreciably larger.
The scalar and pion masses are respectively related to the longitudinal 
and transverse susceptibilities:
\beq
	\chi_{l,t} = Z_{s,\pi} / M^{2}_{s,\pi}
\eeq
The chiral Ward identity~(\ref{eq:chit}) then yields:
\beq 
	M^2_\pi = \frac{Z_\pi}{\langle \bar\chi \chi \rangle} m.
\eeq 
As long as chiral symmetry is broken,
this equation states that the pion mass
squared is a linear function of the bare mass and hence vanishes when
the breaking is dynamical (Goldstone theorem).
Fig.~\ref{fig:pion_vs_bare} shows the behaviour of the pion mass squared
as $m\rightarrow 0$ in the broken phase for $1/g^2=1.8$. The line is a
least squares fit; the fitted intercept lies two standard deviations
from the origin, which is another signal of systematic error.
 
It is satisfying that the spectrum reflects the phase transition so
sharply, moreover in a manner entirely consistent with standard
expectations -- there is no evidence for the unconventional spectrum in
the symmetric phase, associated with an infinite order
phase transition, discussed in~\cite{MirYam96}. Of course, it
would be valuable to refine this initial study both by working closer to
the chiral limit, and with more lattice spacings in the timelike
direction to improve the mass estimates -- clearly at present we are
unable to make reliable estimates of binding energies, and hence
are unable to make detailed comparisons with the predictions of
the $1/N_f$ expansion in the symmetric phase.

\begin{figure}[htbp]
\vspace{3pt}
\centerline{
\setlength\epsfxsize{300pt}
\epsfbox{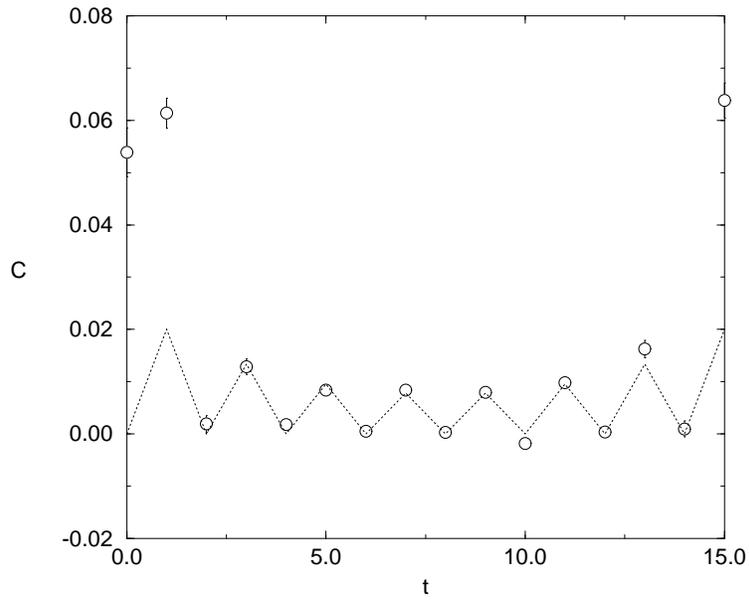}}
\caption{Conserved vector correlator, plus 2 parameter fit, at $1/g^2=2.4$.
\label{fig:vec_corr}}
\end{figure}

\begin{figure}[htbp]
\vspace{3pt}
\centerline{
\setlength\epsfxsize{300pt}
\epsfbox{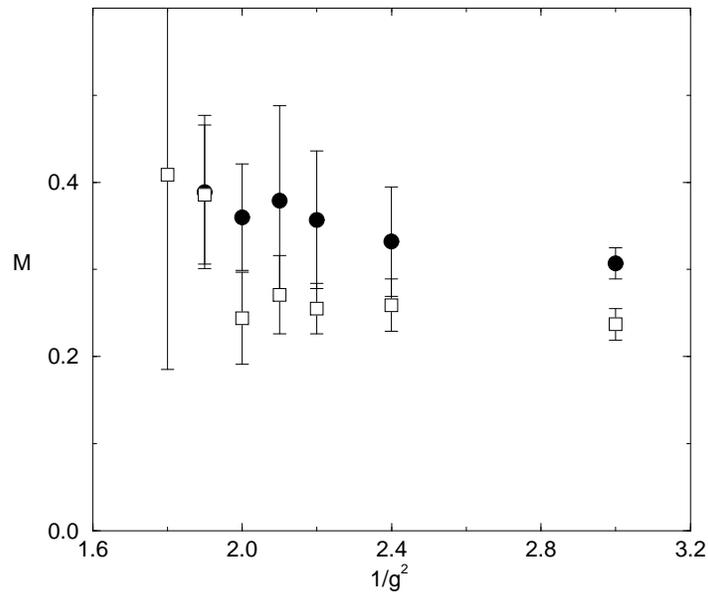}}
\caption{Local (circles) and conserved (squares) 
vector masses from 2 parameter fit vs. $1/g^2$. 
\label{fig:vec_masses}}
\end{figure}

Finally we turn to the vector channels, where we have employed both
local (\ref{eq:Cvl}) and one-link ``conserved'' (\ref{eq:Cvc})
interpolating operators. In either case we found that the signal had a
strong alternating component, exemplified by the plot shown in 
Fig.~\ref{fig:vec_corr}. 
Therefore fits of the form (\ref{eq:altfit}) were used.
Superficially, the main difference between the two different vector
channels was that the local vector gave numerically larger but noisier
signals. For both operators we experimented with two parameter fits,
with $B$ set equal to $-A$ and $M_d$ equal to $M_a$ in (\ref{eq:altfit})
for $t\in[3,13]$, then three parameter fits in which $A$ and $B$ are
allowed to differ, and finally the most general four parameter fit 
in which $A$, $B$, $M_d$ and $M_a$ are all independent, the last two
forms being fitted for $t\in[2,14]$. The results are summarised in 
Tabs. \ref{tab:veclocal}, \ref{tab:veccons}, and plotted in Fig.
~\ref{fig:vec_masses}.

The main trends we found are: 

\noindent{\it(i)\/} For all fits the mass of the local operator lies 
significantly above that of the conserved operator. This should be no
surprise, since as discussed in Sec. 2, and displayed in Tab. 
\ref{tab:spinflav}, the two operators have different quantum numbers in
three dimensions.

\noindent{\it(ii)\/} The masses are approximately constant throughout
the symmetric phase; across the transition the signal becomes smaller
and thus noisier, consistent with the falloff in $\chi_{vns}$ of 
Fig.~\ref{fig:vec_susc}. 
Our mass fits are of limited use for $1/g^2\leq1.8$. In the conserved
channel, however, there is evidence for a sharp rise in mass at
$1/g^2=1.9$. Hence there may be a light vector particle in the symmetric
phase, in accordance with the $1/N_f$ expansion, but not in the broken
phase.

\noindent{\it(iii)\/} Although the two parameter fits gave acceptable
results, there is tentative evidence in the conserved channel that the
four parameter fit is significantly better, with $M_a<M_d$, particularly
from the results with $m=0.02$ at $1/g^2=2.4$. An examination of Tab.
\ref{tab:spinflav} reveals that the direct channel has the vector spin
quantum number form $\bar\psi\gamma_\mu\psi$ while the
alternating channel has the ``axial vector'' form
$\bar\psi\gamma_5\gamma_\mu\psi$. Inspection of the lattice interaction
in the form (\ref{eq:OhGod}) shows that there is a contact interaction
between axial vector currents of equal strength to that between vector
currents; therefore it is reasonable that the one-link operator projects
onto both states with comparable strengths, resulting in $C(t)$ almost
vanishing on even timeslices (Fig.~\ref{fig:vec_corr}). 
Thus we find both light
vectors and light axial vectors in the symmetric phase. We have no
explanation for the light axial vector. Following the analysis 
of~\cite{Hands95}, we can examine the pole condition $p^2=-M_A^2$ in the
axial vector channel to leading order in $1/N_f$:
\bea
1+{g^2\over{2\pi}}\left[m+{{4m^2-M_A^2}\over{4M_A}}\ln\left(
{{2m+M_A}\over{2m-M_A}}\right)\right]=0.
\eea
One readily sees that this equation has no bound state solution of the
form $M_A=2m-\varepsilon$. It would be interesting to see whether 
this persists at next-to-leading order in $1/N_f$.

To summarise, the results in the vector channel present the only
unexpected result; the light state in the axial vector channel is the
first hint of departure from behaviour predicted by the $1/N_f$
expansion in the symmetric phase. In future studies at smaller bare
fermion masses, it would be interesting to monitor the ratio
$\chi_{vs}/\chi_{vns}$, to see whether the channel becomes dominated by
fermion bubble chains as in the $1/N_f$ expansion. It would also be
interesting to repeat this analysis in simulations of non-compact
lattice $\mbox{QED}_3$ to see whether the light axial vector is present
or not -- in this case since the $A_\mu$ field is now a dynamical
variable, it is possible that short-wavelength fluctuations will be
suppressed by a factor $1/k^2$, preventing the strong coupling to the
axial vector channel occurring in (\ref{eq:OhGod}).

\section{Conclusions}

In conclusion we briefly summarise our main findings, and suggest further 
directions to explore. For $N_f=2$ and $N_f=4$ the Thirring model exhibits a 
phase transition
at finite critical coupling $g^2_c$ to a phase in which chiral symmetry is 
spontaneously broken. This contradicts the naive $1/N_f$ expansion, but 
supports the predictions
of non-perturbative methods such as the Schwinger-Dyson approach. 
In the vicinity
of the critical point the model's equation of state seems well fitted by a 
renormalisation-group inspired form corresponding to power-law scaling. 
The picture
remains consistent for $N_f=2$ even once a finite volume scaling analysis, 
using data
from larger lattices and smaller bare masses, is employed. 
The critical exponents 
we find differ significantly between $N_f=2$ and $N_f=4$, 
suggesting that the two
models lie in different universality classes. Although it is tempting to infer 
the existence of  non-perturbative continuum limits at these points, it would 
be nice to have direct evidence by measurement of a renormalised coupling 
constant via a 3- or 4-point correlation function.

For $N_f=6$ the picture is not so clear, and we did not find a fixed-point
fit for 
the equation of state. The Fisher plot of Fig. \ref{fig:Fisher_6} strongly 
suggests
that in the chiral limit the model lies in the symmetric phase for all 
values of the 
coupling, and hence that $N_{fc}<6$. However, bearing in mind the 
practical difficulties 
in identifying the strong coupling limit discussed in Sec. 2, perhaps a 
more conservative
conclusion is that the $N_f=6$ model seems qualitatively very different from
$N_f=2,4$ \cite{Kim96}. 
It might be very difficult ever to exclude absolutely a 
chiral condensate
exponentially suppressed in $N_f$ for large $N_f$; however our data 
disagrees with the only theoretical prediction (\ref{eq:HongPark}) to put 
forward this scenario
\cite{Hong94}. We intend to extend the $N_f=6$ studies to larger lattices and 
smaller $m$ to pursue this issue further.

Although we made no attempt to test the scaling form (\ref{eq:conformal}), 
corresponding
to an infinite order phase transition, our studies of the susceptibilities and
spectrum prejudice us against this scenario: the relative ordering of the 
physical fermion, scalar and pseudoscalar masses
varies continuously across the transition; 
there appear to be light bound states in the 
symmetric phase, comparable with the scale set by the physical fermion mass,
in contradiction to the behaviour expected near a conformal transition discussed
in Refs. \cite{MirYam96,Appel96}. It would be interesting to extend this 
study to
$N_f=4$, where $g_c$ lies closer to the strong coupling limit, to see if the 
spectrum shows any qualitative difference. 
Work in this direction is in progress. 

Finally, although our admittedly exploratory spectrum analysis has 
yielded 
important information, it is clear that larger lattices in the 
timelike direction will be needed before accurate measurements of masses 
and binding energies, 
or indeed departures from the
exponential form of the decay corresponding to a spectral function more 
complicated than 
simple isolated poles, can be accomplished. Accurate mass measurements would be desirable in order to compare with predictions from the $1/N_f$ expansion, 
to check whether
the latter approach has any validity near the transition even in the symmetric 
phase.
There is a first hint of its breaking down in the apparent existence of a light axial vector state here; however it is clear much more work is needed 
both numerically and analytically.

\section{Acknowledgements}

LDD is supported by an EU HCM Institutional Fellowship, 
SJH by a PPARC Advanced Fellowship. 
Some of the computing work was performed using resources 
made available under PPARC research grant GR/J67475,
and those of the UKQCD collaboration under 
GR/K41663, GR/K455745 and GR/L29927. We have enjoyed discussing aspects
of this work with Roger Horsley, Yoonbai Kim, 
Kei-Ichi Kondo and Nick Mavromatos.

\clearpage

\begin{table}[ht]
\setlength{\tabcolsep}{1.5pc}
\caption{List of results for the chiral condensate for $N_f=2$}
\label{tab:Nf=2}
\begin{tabular*}{\textwidth}{@{}l@{\extracolsep{\fill}}rrrrr}
\hline
$L$ &  $m$  & $1/g^2$   & $\cond$ & $\Delta\cond$ \\
\hline
8 & 0.01 & 1.8 & 0.08610 & 0.0030 \\ 
8 & 0.01 & 2.0 & 0.06130 & 0.0020 \\ 
8 & 0.01 & 2.2 & 0.05170 & 0.0020 \\ 
8 & 0.01 & 2.4 & 0.00425 & 0.0010 \\ 
8 & 0.02 & 0.5 & 0.25097 & 0.0084 \\ 
8 & 0.02 & 1.0 & 0.28171 & 0.0110 \\ 
8 & 0.02 & 1.5 & 0.20172 & 0.0068 \\ 
8 & 0.05 & 1.6 & 0.26956 & 0.0080 \\ 
8 & 0.05 & 1.8 & 0.22639 & 0.0060 \\ 
8 & 0.05 & 2.0 & 0.20417 & 0.0040 \\ 
8 & 0.05 & 2.2 & 0.18051 & 0.0043 \\ 
8 & 0.05 & 2.4 & 0.15785 & 0.0038 \\ 
8 & 0.05 & 2.6 & 0.14403 & 0.0035 \\ 
8 & 0.05 & 2.8 & 0.13393 & 0.0028 \\ 
12 & 0.01 & 1.6 & 0.17485 & 0.0057 \\ 
12 & 0.02 & 1.6 & 0.22341 & 0.0055 \\ 
12 & 0.02 & 1.8 & 0.19083 & 0.0047 \\ 
12 & 0.02 & 2.0 & 0.14680 & 0.0040 \\ 
12 & 0.02 & 2.2 & 0.11172 & 0.0026 \\ 
12 & 0.02 & 2.4 & 0.08992 & 0.0020 \\ 
12 & 0.02 & 2.6 & 0.07737 & 0.0015 \\ 
12 & 0.02 & 2.8 & 0.06493 & 0.0010 \\ 
12 & 0.02 & 3.0 & 0.05715 & 0.0008 \\ 
12 & 0.02 & 3.2 & 0.05256 & 0.0007 \\ 
12 & 0.02 & 3.4 & 0.04784 & 0.0006 \\ 
12 & 0.03 & 1.6 & 0.23970 & 0.0035 \\ 
12 & 0.03 & 1.8 & 0.20873 & 0.0036 \\ 
12 & 0.03 & 2.0 & 0.17070 & 0.0023 \\ 
12 & 0.03 & 2.2 & 0.14342 & 0.0025 \\ 
12 & 0.03 & 2.4 & 0.12171 & 0.0017 \\ 
12 & 0.03 & 2.6 & 0.10508 & 0.0013 \\ 
12 & 0.03 & 2.8 & 0.09153 & 0.0011 \\ 
12 & 0.03 & 3.0 & 0.08303 & 0.0008 \\ 
12 & 0.03 & 3.2 & 0.07554 & 0.0008 \\ 
12 & 0.03 & 3.4 & 0.07107 & 0.0007 \\ 
\hline
\end{tabular*}
\end{table}

\begin{table}[ht]
\setlength{\tabcolsep}{1.5pc}
\caption{List of results for $N_f=2$ (continued)}
\label{tab:Nf=2_bis}
\begin{tabular*}{\textwidth}{@{}l@{\extracolsep{\fill}}rrrrr}
\hline
$L$ &  $m$  & $1/g^2$   & $\cond$ & $\Delta\cond$ \\
\hline
12 & 0.04 & 1.6 & 0.25599 & 0.0031 \\ 
12 & 0.04 & 1.8 & 0.22321 & 0.0032 \\ 
12 & 0.04 & 2.0 & 0.19460 & 0.0030 \\ 
12 & 0.04 & 2.2 & 0.17244 & 0.0024 \\ 
12 & 0.04 & 2.4 & 0.14307 & 0.0019 \\ 
12 & 0.04 & 2.6 & 0.12966 & 0.0018 \\ 
12 & 0.04 & 2.8 & 0.11604 & 0.0015 \\ 
12 & 0.04 & 3.0 & 0.10851 & 0.0012 \\ 
12 & 0.04 & 3.2 & 0.09707 & 0.0011 \\ 
12 & 0.04 & 3.4 & 0.09282 & 0.0008 \\ 
12 & 0.05 & 1.6 & 0.27098 & 0.0030 \\ 
12 & 0.05 & 1.8 & 0.23624 & 0.0028 \\ 
12 & 0.05 & 2.0 & 0.20952 & 0.0025 \\ 
12 & 0.05 & 2.2 & 0.18450 & 0.0020 \\ 
12 & 0.05 & 2.4 & 0.16750 & 0.0020 \\ 
12 & 0.05 & 2.6 & 0.15171 & 0.0020 \\ 
12 & 0.05 & 2.8 & 0.13861 & 0.0014 \\ 
12 & 0.05 & 3.0 & 0.12572 & 0.0012 \\ 
12 & 0.05 & 3.2 & 0.11861 & 0.0011 \\ 
12 & 0.05 & 3.4 & 0.11118 & 0.0010 \\ 
16 & 0.01 & 1.6 & 0.20250 & 0.0028 \\ 
16 & 0.01 & 1.8 & 0.14190 & 0.0021 \\ 
16 & 0.01 & 1.9 & 0.12780 & 0.0032 \\ 
16 & 0.01 & 1.92 & 0.12419 & 0.0018 \\ 
16 & 0.01 & 2.0 & 0.10250 & 0.0021 \\ 
16 & 0.01 & 2.1 & 0.08614 & 0.0020 \\ 
16 & 0.01 & 2.2 & 0.07177 & 0.0012 \\ 
16 & 0.01 & 2.4 & 0.05287 & 0.0010 \\ 
16 & 0.02 & 1.6 & 0.22040 & 0.0030 \\ 
16 & 0.02 & 1.92 & 0.15733 & 0.0013 \\ 
16 & 0.02 & 2.0 & 0.14580 & 0.0012 \\ 
16 & 0.02 & 2.4 & 0.09430 & 0.0006 \\ 
16 & 0.03 & 1.8 & 0.20440 & 0.0010 \\ 
16 & 0.03 & 1.9 & 0.18624 & 0.0009 \\ 
16 & 0.03 & 1.92 & 0.18547 & 0.0010 \\ 
16 & 0.03 & 2.0 & 0.17174 & 0.0009 \\ 
16 & 0.03 & 2.1 & 0.15632 & 0.0009 \\ 
16 & 0.04 & 1.92 & 0.20576 & 0.0008 \\ 

\hline
\end{tabular*}
\end{table}

\begin{table}[ht]
\setlength{\tabcolsep}{1.5pc}
\caption{List of results for $N_f=4$}
\label{tab:Nf=4}
\begin{tabular*}{\textwidth}{@{}l@{\extracolsep{\fill}}rrrrr}
\hline
$L$ &  $m$  & $1/g^2$   & $\cond$ & $\Delta\cond$ \\
\hline
12 & 0.02 & 0.6 & 0.21771 & 0.0037 \\ 
12 & 0.02 & 0.7 & 0.17435 & 0.0032 \\ 
12 & 0.02 & 0.8 & 0.13163 & 0.0030 \\ 
12 & 0.02 & 0.9 & 0.09994 & 0.0024 \\ 
12 & 0.02 & 1.0 & 0.07917 & 0.0018 \\ 
12 & 0.02 & 1.1 & 0.06700 & 0.0011 \\ 
12 & 0.02 & 1.2 & 0.06030 & 0.0008 \\ 
12 & 0.02 & 1.3 & 0.05453 & 0.0007 \\ 
12 & 0.02 & 1.4 & 0.04882 & 0.0005 \\ 
12 & 0.02 & 1.5 & 0.04636 & 0.0005 \\ 
12 & 0.03 & 0.6 & 0.23476 & 0.0033 \\ 
12 & 0.03 & 0.7 & 0.20343 & 0.0031 \\ 
12 & 0.03 & 0.8 & 0.16221 & 0.0028 \\ 
12 & 0.03 & 0.9 & 0.13419 & 0.0020 \\ 
12 & 0.03 & 1.0 & 0.11065 & 0.0017 \\ 
12 & 0.03 & 1.1 & 0.09750 & 0.0020 \\ 
12 & 0.03 & 1.2 & 0.08977 & 0.0016 \\
12 & 0.03 & 1.3 & 0.07804 & 0.0009 \\ 
12 & 0.03 & 1.4 & 0.07295 & 0.0009 \\ 
12 & 0.03 & 1.5 & 0.06793 & 0.0006 \\ 
12 & 0.04 & 0.6 & 0.24741 & 0.0036 \\ 
12 & 0.04 & 0.7 & 0.21080 & 0.0028 \\ 
12 & 0.04 & 0.8 & 0.18680 & 0.0024 \\ 
12 & 0.04 & 0.9 & 0.16079 & 0.0030 \\ 
12 & 0.04 & 1.0 & 0.13808 & 0.0020 \\ 
12 & 0.04 & 1.1 & 0.12140 & 0.0018 \\ 
12 & 0.04 & 1.2 & 0.10850 & 0.0013 \\ 
12 & 0.04 & 1.3 & 0.10094 & 0.0010 \\ 
12 & 0.04 & 1.4 & 0.09362 & 0.0009 \\ 
12 & 0.04 & 1.5 & 0.08734 & 0.0008 \\ 
12 & 0.05 & 0.6 & 0.25489 & 0.0026 \\ 
12 & 0.05 & 0.7 & 0.23618 & 0.0025 \\ 
12 & 0.05 & 0.8 & 0.20854 & 0.0022 \\ 
12 & 0.05 & 0.9 & 0.18000 & 0.0021 \\ 
12 & 0.05 & 1.0 & 0.160 & 0.0016 \\ 
12 & 0.05 & 1.1 & 0.14423 & 0.0015 \\ 
12 & 0.05 & 1.2 & 0.12921 & 0.0013 \\ 
12 & 0.05 & 1.3 & 0.12051 & 0.0012 \\ 
12 & 0.05 & 1.4 & 0.11202 & 0.0011 \\ 
12 & 0.05 & 1.5 & 0.10653 & 0.0010 \\ 
\hline
\end{tabular*}
\end{table}

\begin{table}[ht]
\setlength{\tabcolsep}{1.5pc}
\caption{List of results for $N_f=6$}
\label{tab:Nf=6}
\begin{tabular*}{\textwidth}{@{}l@{\extracolsep{\fill}}rrrrr}
\hline
$L$ &  $m$  & $1/g^2$   & $\cond$ & $\Delta\cond$ \\
\hline
12 & 0.02 & 0.25 & 0.14342 & 0.0034 \\
12 & 0.02 & 0.3 & 0.14687 & 0.0035 \\ 
12 & 0.02 & 0.35 & 0.13924 & 0.0040 \\ 
12 & 0.02 & 0.4 & 0.11832 & 0.0030 \\ 
12 & 0.02 & 0.45 & 0.09867 & 0.0040 \\ 
12 & 0.02 & 0.5 & 0.07608 & 0.0020 \\ 
12 & 0.03 & 0.25 & 0.16721 & 0.0030 \\ 
12 & 0.03 & 0.3 & 0.17836 & 0.0040 \\ 
12 & 0.03 & 0.35 & 0.17099 & 0.0032 \\ 
12 & 0.03 & 0.4 & 0.15218 & 0.0028 \\ 
12 & 0.03 & 0.45 & 0.12766 & 0.0030 \\ 
12 & 0.03 & 0.5 & 0.10716 & 0.0020 \\ 
12 & 0.04 & 0.3 & 0.18403 & 0.0029 \\ 
12 & 0.04 & 0.35 & 0.18659 & 0.0031 \\ 
12 & 0.04 & 0.4 & 0.17549 & 0.0026 \\ 
12 & 0.04 & 0.45 & 0.15650 & 0.0022 \\ 
12 & 0.04 & 0.5 & 0.13959 & 0.0038 \\ 
12 & 0.05 & 0.25 & 0.19691 & 0.0028 \\ 
12 & 0.05 & 0.3 & 0.20373 & 0.0029 \\ 
12 & 0.05 & 0.35 & 0.20636 & 0.0028 \\ 
12 & 0.05 & 0.4 & 0.19784 & 0.0030 \\ 
12 & 0.05 & 0.45 & 0.17549 & 0.0017 \\ 
12 & 0.05 & 0.5 & 0.15903 & 0.0020 \\ 

\hline
\end{tabular*}
\end{table}

\begin{table}[ht]
\setlength{\tabcolsep}{1.5pc}
\caption{Spin/flavor assignments of the fermion bilinears used in
spectroscopy}
\label{tab:spinflav}
\begin{tabular*}{\textwidth}{@{}l@{\extracolsep{\fill}}rrrrrr}
\hline
& direct & alternating \\
\hline
pion    & 
$\gamma_5\otimes\One$ &  $\gamma_\mu\gamma_\nu\otimes\tau_\mu^*
\tau_\nu^*$  \\
scalar  & 
$\One\otimes\One$ & $\gamma_5\gamma_\mu\gamma_\nu\otimes\tau_\mu^*
\tau_\nu^*$  \\
local vector  & 
$\gamma_\mu\gamma_3\otimes\tau_3^*\tau_\mu^*$ & 
$\gamma_5\gamma_\mu\gamma_3\otimes\tau_3^*
\tau_\mu^*$  \\
conserved vector  & 
$\gamma_\mu\otimes\One$ & 
$\gamma_5\gamma_\mu\otimes\tau_3^*$ \\
\hline
\end{tabular*}
\end{table}

\begin{table}[ht]
\setlength{\tabcolsep}{1.5pc}
\caption{Results from fits on the $12^3$ lattice.}
\label{tab:fit1}
\begin{tabular*}{\textwidth}{@{}l@{\extracolsep{\fill}}rrrr}
\hline
        & Parameter      &   Fit I     &   Fit II     \\
\hline
$N_f=2$ & $1/g^2_c$      &  2.03(9)    &  1.94(4)     \\
        & $\delta$       &  2.32(23)   &  2.68(16)    \\
        & $\beta$        &  0.71(9)    &  ---         \\
        & $A$            &  0.32(5)    &  0.37(1)     \\
        & $B$            &  1.91(43)   &  2.86(35)    \\
        & $\chi^2$/d.o.f &  2.4        &  2.1         \\
        &                &             &              \\
$N_f=4$ & $1/g^2_c$      &  0.63(1)    &  0.66(1)     \\
        & $\delta$       &  3.67(28)   &  3.43(19)    \\
        & $\beta$        &  0.38(4)    &  ---     \\
        & $A$            &  0.78(5)    &  0.73(2)     \\
        & $B$            &  7.9(2.8)   &  6.4(1.5)    \\
        & $\chi^2$/d.o.f &  3.1        &  2.0         \\
\hline
\end{tabular*}
\end{table}

\begin{table}[ht]
\caption{Results from fits on the $16^3$ lattice.}
\label{tab:fit2}
\begin{tabular*}{\textwidth}{@{}l@{\extracolsep{\fill}}rrrr}
\hline
        & Parameter      &   Fit I     &   Fit II     \\
\hline
$N_f=2$ & $1/g^2_c$      &             &  1.93(4)     \\
        & $\delta$       &             &  2.55(15)    \\
        & $\beta$        &             &  ---         \\
        & $A$            &             &  0.38(1)     \\
        & $B$            &             &  2.29(35)    \\
        & $\chi^2$/d.o.f &             &  2.3         \\
\hline
\end{tabular*}
\end{table}

\begin{table}[ht]
\caption{Results from fit including finite size scaling.}
\label{tab:fit3}
\begin{minipage}{\linewidth}
\renewcommand{\thefootnote}{\thempfootnote}
\begin{tabular*}{\textwidth}{@{}l@{\extracolsep{\fill}}rrr}
\hline
        & Parameter      &   Fit III   \\
\hline
$N_f=2$ & $1/g^2_c$      &  1.92(2)    \\
        & $\delta$       &  2.75(9)    \\
        & $\beta$~\footnote{evaluated from 
$\delta-1/\beta=1$ constraint}
                         &  0.57(2)    \\
        & $\eta$~\footnote{evaluated from hyperscaling relation}
                         &  0.60(2)    \\
        & $\nu$~\footnotemark[\value{mpfootnote}]
                         &  0.71(4)    \\
        & $A$            &  0.334(7)   \\
        & $B$            &  2.7(3)     \\
        & $C$            &  2.1(7)     \\
        & $\chi^2$/d.o.f &  1.76       \\
\hline
\end{tabular*}
\end{minipage}
\end{table}

\begin{table}[ht]
\caption{Longitudinal and transverse susceptibilities in the vicinity of
the critical coupling.}
\label{tab:R}
\bigskip
\begin{minipage}{\linewidth}
\renewcommand{\thefootnote}{\thempfootnote}
\begin{tabular*}{\textwidth}{@{}r@{\extracolsep{\fill}}lllllll}
\hline
$1/g^2$ & $m$ & $\chi_{ls}$ & $\chi_{lns}$ & $\chi_l$ & $\chi_t$ & 
$R=\chi_l/\chi_t$ \\
\hline
1.9 & 0.01  & 1.51(15) & 3.40(40) & 4.90(43) & 12.49(20) & 0.392(35) \\
    & 0.03  & 0.57(8)  & 1.68(15) & 2.24(17) &  6.22(4)  & 0.361(28) \\
\hline
1.92 & 0.01 & 1.65(19) & 3.88(28) & 5.53(34) & 12.17(25) & 0.454(29) \\
     & 0.02 & 0.72(8)  & 2.41(15) & 3.13(17) & 7.83(7)   & 0.400(22) \\
     & 0.03 & 0.67(9)  & 1.70(10) & 2.37(13) & 6.17(5)   & 0.384(21) \\
     & 0.04 & 0.34(5)  & 1.35(8)  & 1.69(9)  & 5.15(3)   & 0.329(18) \\
\hline
2.0  & 0.01 & 1.13(19) & 4.25(44) & 5.37(48) & 10.12(21) & 0.531(49) \\
     & 0.03 & 0.62(8)  & 1.93(9)  & 2.54(12) & 5.71(5)   & 0.445(21) \\
\hline
\end{tabular*}
\end{minipage}
\end{table}
\begin{table}[ht]
\caption{Masses for the fermion, pion and scalar, $16^3$ lattice, $m=0.01$}
\label{tab:16_3_001_mass}
\bigskip
\begin{minipage}{\linewidth}
\renewcommand{\thefootnote}{\thempfootnote}
\begin{tabular*}{\textwidth}{@{}r@{\extracolsep{\fill}}llll}
\hline
$1/g^2$ & $\mu_R$ & $M_\pi$ & $M_s$ \\
\hline
1.6  & 0.36 (3) & 0.20 (1) & ---      \\
1.8  & 0.23 (2) & 0.23 (2) & 0.47 (4) \\
1.9  & 0.16 (1) & 0.21 (1) & 0.33 (2) \\
2.0  & 0.10 (1) & 0.24 (1) & 0.29 (1) \\
2.1  & 0.10 (1) & 0.22 (1) & 0.27 (1) \\
2.2  & 0.08 (1) & 0.21 (1) & 0.26 (1) \\
2.4  & 0.06 (1) & 0.21 (1) & 0.23 (1) \\
\hline
\end{tabular*}
\end{minipage}
\end{table}

\begin{table}[ht]
\caption{Masses for the fermion, pion and scalar, $16^3$ lattice, $m=0.02$}
\label{tab:16_3_002_mass}
\bigskip
\begin{minipage}{\linewidth}
\renewcommand{\thefootnote}{\thempfootnote}
\begin{tabular*}{\textwidth}{@{}r@{\extracolsep{\fill}}llll}
\hline
$1/g^2$ & $\mu_R$ & $M_\pi$ & $M_s$ \\
\hline
1.6  & 0.330 (40) & 0.280 (4) & 0.52 (15) \\
1.8  & 0.360 (40) & 0.290 (6) & 0.39  (8) \\
1.92 & 0.200 (10) & 0.280 (4) & 0.43  (6) \\
2.0  & 0.190 (10) & 0.300 (5) & 0.35  (5) \\
2.4  & 0.109  (3) & 0.290 (4) & 0.27  (1) \\
\hline
\end{tabular*}
\end{minipage}
\end{table}

\begin{table}[ht]
\caption{Masses for the fermion, pion and scalar, $16^3$ lattice, $m=0.03$}
\label{tab:16_3_003_mass}
\bigskip
\begin{minipage}{\linewidth}
\renewcommand{\thefootnote}{\thempfootnote}
\begin{tabular*}{\textwidth}{@{}r@{\extracolsep{\fill}}llll}
\hline
$1/g^2$ & $\mu_R$ & $M_\pi$ & $M_s$ \\
\hline
1.8  & 0.290 (20) & 0.340 (5) & 0.64 (9) \\
1.9  & 0.240 (20) & 0.338 (4) & 0.60 (4) \\
1.92 & 0.250 (20) & 0.338 (4) & 0.64 (6) \\
2.0  & 0.270 (20) & 0.344 (4) & 0.54 (4) \\
2.1  & 0.210 (10) & 0.341 (4) & 0.50 (3) \\
\hline
\end{tabular*}
\end{minipage}
\end{table}

\begin{table}[ht]
\caption{Masses for the fermion, pion and scalar, $8^2\times 16$ lattice, 
$m=0.02$}
\label{tab:8_2_16_002_mass}
\bigskip
\begin{minipage}{\linewidth}
\renewcommand{\thefootnote}{\thempfootnote}
\begin{tabular*}{\textwidth}{@{}r@{\extracolsep{\fill}}llll}
\hline
$1/g^2$ & $\mu_R$ & $M_\pi$ & $M_s$ \\
\hline
1.6  &    ****    &   ****    & **** \\
1.7  & 0.340 (30) & 0.256 (4) & 0.42 (5) \\
1.9  & 0.290 (20) & 0.240 (5) & 0.44 (7) \\
2.0  & 0.280 (10) & 0.231 (6) & 0.30 (7) \\
2.1  & 0.200 (10) & 0.231 (6) & 0.26 (3) \\
2.2  & 0.240 (10) & 0.225 (5) & 0.29 (2) \\
2.3  & 0.185  (5) & 0.225 (6) & 0.23 (2) \\
2.4  & 0.170  (9) & 0.206 (5) & 0.26 (2) \\
\hline
\end{tabular*}
\end{minipage}
\end{table}

\begin{table}[ht]
\caption{Masses for the fermion, pion and scalar, $8^2\times 16$ lattice, 
$m=0.05$}
\label{tab:8_2_16_005_mass}
\bigskip
\begin{minipage}{\linewidth}
\renewcommand{\thefootnote}{\thempfootnote}
\begin{tabular*}{\textwidth}{@{}r@{\extracolsep{\fill}}llll}
\hline
$1/g^2$ & $\mu_R$ & $M_\pi$ & $M_s$ \\
\hline
1.6  & 0.480 (40) & 0.398 (4) & --- \\
1.7  & 0.470 (30) & 0.390 (4) & --- \\
1.8  &    ---     & 0.390 (4) & --- \\
1.9  & 0.350 (10) & 0.386 (3) & --- \\
2.0  & 0.350 (20) & 0.378 (4) & --- \\
2.1  & 0.320 (10) & 0.368 (4) & --- \\
2.2  & 0.280 (10) & 0.362 (4) & --- \\
2.3  &    ---     & 0.359 (3) & --- \\
2.4  & 0.255  (9) & 0.349 (3) & --- \\
2.5  & 0.246  (7) & 0.340 (3) & --- \\ 
\hline
\end{tabular*}
\end{minipage}
\end{table}

\begin{table}[ht]
\caption{Masses obtained in the local vector channel from a $16^3$ lattice}
\label{tab:veclocal}
\bigskip
\begin{minipage}{\linewidth}
\renewcommand{\thefootnote}{\thempfootnote}
\begin{tabular*}{\textwidth}{@{}r@{\extracolsep{\fill}}lllll}
\hline
\# fitted parameters& $m$ & $1/g^2$ & $M_d$ & $M_a$ & $\chi^2$/dof \\
\hline
2 & 0.01 & 3.0 & 0.307(18) & & 7.1\\
  &      & 2.4 & 0.332(63) & & 1.4\\
  &      & 2.2 & 0.357(79) & & 4.2\\
  &      & 2.1 & 0.379(109) & & 1.2\\
  &      & 2.0 & 0.360(61) & & 1.7\\
  &      & 1.9 & 0.389(88) & & 0.4\\
  &      & 1.8 & 0.807(665) & & 2.3\\
\hline
  & 0.02 & 2.4 & 0.365(34) & & 4.2\\
\hline
  & 0.03 & 2.1 & 0.952(183) & & 3.7\\
  &      & 2.0 & 0.870(171) & & 2.6\\
  &      & 1.9 & 0.438(88)  & & 2.5\\
  &      & 1.8 & 0.928(333) & & 1.5\\
\hline
\hline
3 & 0.01 & 3.0 & 0.310(19) & & 3.4\\
  &      & 2.4 & 0.350(57) & & 0.8\\
  &      & 2.2 & 0.388(67) & & 1.8\\
  &      & 2.1 & 0.479(132)& & 1.5\\
  &      & 2.0 & 0.358(59) & & 1.0\\
  &      & 1.9 & 0.397(87) & & 0.3\\
  &      & 1.8 & 0.764(495)& & 2.4\\
\hline
\hline
4 & 0.01 & 3.0 & 0.263(18) & 0.256(19) & 1.2\\
  &      & 2.4 & 0.309(57) & 0.294(62) & 0.7\\
  &      & 2.2 & 0.336(66) & 0.284(100)& 1.9\\
  &      & 2.1 & 0.448(106)& 0.312(103)& 1.1\\
  &      & 2.0 & 0.358(60) & 0.345(79) & 1.3\\
  &      & 1.9 & 0.395(100)& 0.337(173)& 0.5\\
  &      & 1.8 & 0.701(457)& 1.368(2.7)& 2.5\\
\hline
\end{tabular*}
\end{minipage}
\end{table}

\begin{table}[ht]
\caption{Masses obtained in the conserved vector channel from a $16^3$ lattice}
\label{tab:veccons}
\bigskip
\begin{minipage}{\linewidth}
\renewcommand{\thefootnote}{\thempfootnote}
\begin{tabular*}{\textwidth}{@{}r@{\extracolsep{\fill}}lllll}
\hline
\# fitted parameters& $m$ & $1/g^2$ & $M_d$ & $M_a$ & $\chi^2$/dof \\
\hline
2 & 0.01 & 3.0 & 0.237(18) & & 3.0\\
  &      & 2.4 & 0.259(30) & & 2.1\\
  &      & 2.2 & 0.255(29) & & 1.3\\
  &      & 2.1 & 0.271(45) & & 0.7\\
  &      & 2.0 & 0.244(53) & & 1.5\\
  &      & 1.9 & 0.386(80) & & 0.5\\
  &      & 1.8 & 0.409(224) & & 0.6\\
\hline
  & 0.02 & 2.4 & 0.306(18) & & 2.2\\
\hline
  & 0.03 & 2.1 & 0.489(45) & & 1.3\\
  &      & 2.0 & 0.656(75) & & 1.9\\
  &      & 1.9 & 0.566(114)  & & 2.8\\
  &      & 1.8 & 0.814(218) & & 0.9\\
\hline
\hline
3 & 0.01 & 3.0 & 0.239(15) & & 2.8\\
  &      & 2.4 & 0.266(30) & & 1.9\\
  &      & 2.2 & 0.257(29) & & 1.5\\
  &      & 2.1 & 0.312(42)& & 0.5\\
  &      & 2.0 & 0.258(56) & & 1.8\\
  &      & 1.9 & 0.456(89) & & 0.9\\
  &      & 1.8 & 0.974(879)& & 0.9\\
\hline
\hline
4 & 0.01 & 3.0 & 0.253(18) & 0.230(17) & 2.9\\
  &      & 2.4 & 0.293(34) & 0.229(36) & 1.8\\
  &      & 2.2 & 0.325(37) & 0.162(46)& 0.5\\
  &      & 2.1 & 0.320(43)& 0.259(73)& 0.5\\
  &      & 2.0 & 0.302(63) & 0.223(67) & 1.9\\
  &      & 1.9 & 0.605(204)& 0.190(203)& 0.5\\
\hline
  & 0.02 & 2.4 & 0.366(20) & 0.220(23) & 1.2\\
\hline
\end{tabular*}
\end{minipage}
\end{table}

\clearpage

\vskip 2 truecm

\vfill\eject

\end{document}